\documentclass{aa}

\usepackage{graphics}
\usepackage{graphicx}
\usepackage{multicol}
\usepackage{longtable}
\usepackage{supertabular}
\usepackage{txfonts}

\usepackage{natbib,twoopt}
\usepackage{amsmath} 
\usepackage{hyperref} 
\bibpunct{(}{)}{;}{a}{}{,} 
\makeatletter
\newcommandtwoopt{\citeads}[3][][]{\href{http://adsabs.harvard.edu/abs/#3}%
{\def\hyper@linkstart##1##2{}%
\let\hyper@linkend\@empty\citealp[#1][#2]{#3}}}
\newcommandtwoopt{\citepads}[3][][]{\href{http://adsabs.harvard.edu/abs/#3}%
{\def\hyper@linkstart##1##2{}%
\let\hyper@linkend\@empty\citep[#1][#2]{#3}}}
\newcommandtwoopt{\citetads}[3][][]{\href{http://adsabs.harvard.edu/abs/#3}%
{\def\hyper@linkstart##1##2{}%
\let\hyper@linkend\@empty\citet[#1][#2]{#3}}}
\newcommandtwoopt{\citeyearads}[3][][]%
{\href{http://adsabs.harvard.edu/abs/#3}
{\def\hyper@linkstart##1##2{}%
\let\hyper@linkend\@empty\citeyear[#1][#2]{#3}}}
\makeatother

\usepackage{color}
\definecolor{mygreen}{RGB}{0,128,0}

\hypersetup{colorlinks=true,linkcolor=blue,citecolor=blue,urlcolor=blue}

\def\NtotalpmanomalyCEP{254}
\def\NlowpmanomalyCEP{75}
\def\NmidpmanomalyCEP{26}
\def\NhighpmanomalyCEP{31}
\def\pmanomalyfractionCEP{22}

\def\NtotalpmanomalyRRL{189}
\def\NlowpmanomalyRRL{61}
\def\NmidpmanomalyRRL{8}

\begin{document}

\title{Multiplicity of Galactic Cepheids and RR Lyrae stars from Gaia DR2}
\subtitle{I. Binarity from proper motion anomaly\thanks{Tables \ref{pm_binaries1} and \ref{pm_rrlyr_1}
are available in electronic form at the CDS via anonymous ftp to \url{cdsarc.u-strasbg.fr} (130.79.128.5) or via \url{http://cdsweb.u-strasbg.fr/cgi-bin/qcat?J/A+A/}}}
\titlerunning{Multiplicity of Galactic Cepheids and RR Lyrae stars from Gaia DR2 - I}
\authorrunning{P. Kervella et al.}
\author{
Pierre~Kervella\inst{1}
\and
Alexandre Gallenne\inst{2}
\and
Nancy Remage Evans\inst{3}
\and
Laszlo Szabados\inst{4}
\and
Fr\'ed\'eric Arenou\inst{5}
\and
Antoine M\'erand\inst{6}
\and
Yann Proto\inst{7,1}
\and
Paulina Karczmarek\inst{8}
\and
Nicolas Nardetto\inst{9}
\and
Wolfgang Gieren\inst{10}
\and
Grzegorz Pietrzynski\inst{11}
}
\institute{
LESIA, Observatoire de Paris, Universit\'e PSL, CNRS, Sorbonne Universit\'e, Univ. Paris Diderot, Sorbonne Paris Cit\'e, 5 place Jules Janssen, 92195 Meudon, France, \email{pierre.kervella@obspm.fr}.
\and
European Southern Observatory, Alonso de C\'ordova 3107, Casilla 19001, Santiago, Chile.
\and
Smithsonian Astrophysical Observatory, MS 4, 60 Garden Street, Cambridge, MA 02138, USA.
\and
Konkoly Observatory, MTA CSFK, Konkoly Thege M. \'ut 15-17, H-1121, Hungary.
\and
GEPI, Observatoire de Paris, Universit\'e PSL, CNRS, 5 Place Jules Janssen, 92190 Meudon, France.
\and
European Southern Observatory, Karl-Schwarzschild-Str. 2, 85748 Garching, Germany.
\and
Ecole Normale Sup\'erieure de Lyon, 15 parvis Ren\'e Descartes, 69342 Lyon, France.
\and
Warsaw University Observatory, Al. Ujazdowskie 4, 00-478, Warsaw, Poland.
\and
Universit\'e C\^ote d'Azur, OCA, CNRS, Lagrange, France
\and
Universidad de Concepci{\'o}n, Departamento de Astronom\'{\i}a, Casilla 160-C, Concepci{\'o}n, Chile.
\and
Nicolaus Copernicus Astronomical Centre, Polish Academy of Sciences, Bartycka 18, PL-00-716 Warszawa, Poland
}
\date{Received ; Accepted}
\abstract {Classical Cepheids (CCs) and RR Lyrae stars (RRLs) are important classes of variable stars used as standard candles to estimate galactic and extragalactic distances. Their multiplicity is imperfectly known, particularly for RRLs. Astoundingly, to date only one RRL has convincingly been demonstrated to be a binary, TU UMa, out of tens of thousands of known RRLs.}
{Our  aim is to detect the binary and multiple stars present in a sample of Milky Way CCs and RRLs.}
{In the present article, we combine the Hipparcos and Gaia DR2 positions to determine the mean proper motion of the targets, and we search for proper motion anomalies (PMa) caused by close-in orbiting companions.}
{We identify 57 CC binaries from PMa out of 254 tested stars and 75 additional candidates, confirming the high binary fraction of these massive stars. For 28 binary CCs, we determine the companion mass by combining their spectroscopic orbital parameters and astrometric PMa.
We detect 13 RRLs showing a significant PMa out of 198 tested stars, and 61 additional candidates.} 
{We determine that the binary fraction of CCs is likely above 80\%, while that of RRLs is at least 7\%. The newly detected systems will be useful to improve our understanding of their evolutionary states. The discovery of a significant number of RRLs in binary systems also resolves the long-standing mystery of their extremely low apparent binary fraction.}

\maketitle


\section{Introduction}

The remarkable correlation of the intrinsic luminosity of classical Cepheids (CCs) \citepads{1908AnHar..60...87L,1912HarCi.173....1L,2007A&A...476...73F} and RR Lyrae stars (RRLs) \citepads{2004ApJS..154..633C,2017ApJ...841...84N} respectively with their pulsation period and metallicity makes these two classes of variable stars essential standard candles for Galactic \citepads{2013ApJ...763...32D}, globular cluster \citepads{1992ApJ...386..663C}, and extragalactic distance measurements \citepads{2003AJ....125.1309C,2011ApJ...730..119R,2016ApJ...826...56R}.
An analysis of the GDR2 parallaxes of Galactic CCs in the context of the extragalactic distance scale was recently presented by \citetads{2018ApJ...861..126R}.
Classical Cepheids are intermediate-mass stars (typically 5 to $10\,M_\odot$) and this class therefore comprises a high fraction of binary and multiple stars \citepads{1982ApJS...49....1G,2003ASPC..298..237S,2013AJ....146...93E,2015AJ....150...13E,2017IAUS..329..110S}.
RR Lyrae stars  are short-period ($P\approx 0.5$\,d) low-mass pulsators ($m \approx 0.6\,M_\odot$) that are abundant in the Galaxy and especially in globular clusters \citepads{1899ApJ....10..255B, 1901ApJ....13..226P}.
They are old (age $\approx 10$\,Ga\footnote{``a'' is the recommended IAU symbol for the Julian year as per the 1989 IAU Style Manual \citepads{1990IAUTB..20S....W}, summarized at \url{https://www.iau.org/publications/proceedings_rules/units/}.}), horizontal branch stars, with a typical radius of 4 to $8\,R_\odot$ \citepads{1538-4357-623-2-L133}.
Despite their relative faintness compared to CCs, RRLs have been used to measure distances in the Local Group and beyond \citepads{2010ApJ...708L.121D, 2017SSRv..212.1743D,2018MNRAS.479.4279M}, and their ubiquity makes them important standard candles for Galactic astronomy \citepads{2018ApJ...857...54D, 2018arXiv180704303C}.
Approximately two hundred thousand variable stars are classified as RRLs from ground-based surveys or the Gaia DR2 (\citeads{2014AcA....64..177S}, \citeads{2018arXiv180502079C}, \citeads{2018MNRAS.481.1195M}, \citeads{2018A&A...618A..30H}, \citeads{2018arXiv181103919R}).
It is  remarkable, however,  that there is to date very little evidence for RRLs in binary systems, as only one case has been convincingly identified: \object{TU UMa} \citepads{2016A&A...589A..94L}.
From the analysis of the light curves of a large sample of nearly 2000 RR Lyrae stars, \citetads{2015MNRAS.449L.113H} identified 12 binary candidates that display phase shifts of their light curves that can be attributed to light-time effect (LiTE) that point to the presence of an orbiting companion.
\citetads{2016MNRAS.459.4360L} presents a study of 11 systems searching for LiTE in O-C diagrams, and in the \texttt{RRLyrBinCan}\footnote{\url{http://rrlyrbincan.physics.muni.cz}} database of candidate binary RRLs.
A search has also been conducted by \citetads{2015EPJWC.10106030G} in the Kepler mission light curve database, but without detection. \citetads{2017MNRAS.465L...1S} has interpreted the Kepler light curve phase modulations of \object{KIC 2831097} as being caused by the presence of an orbiting black hole of $8.4\,M_\odot$, but the radial velocity data contradict the binary interpretation.
The common presence of the Blazhko effect \citepads{1907AN....175..325B, 2011MNRAS.411.1763J, 2012MNRAS.424.3094S, 2017MNRAS.470..617J} and period drifts \citepads{2011MNRAS.411.1744S} in many RRLs complicates the uniqueness of the interpretation of the observed phase shifts.
\citetads{2011AcA....61....1S} found a likely candidate for an RRL in a 15.2-day period eclipsing binary system (OGLE-BLG-RRLYR-02792). It was subsequently identified as a peculiar type of ``RR Lyr impostor'' \citepads{2012Natur.484...75P, 2013MNRAS.428.3034S}, and was included in a new class of binary evolution pulsators (BEP) that are believed to be very rare \citepads{2017MNRAS.466.2842K}.

The companions of CCs and RRLs are important for several reasons \citepads{2010EAS....45..441S}.
They may influence their evolution through mass transfer.
They also shift the apparent brightness of their parent stars in a systematically positive way by up to 10\% or more in the visible \citepads{2013A&A...552A..21G}, which affects the zero point of the CC Leavitt law.
\citetads{2016ApJS..226...18A} showed that companions have a limited effect on the observational properties of the brighter, long-period CCs, whose usefulness as distance indicators is thus not affected (see also \citeads{2018ApJ...861...36A}).
However, due to their lower intrinsic brightness, short- and intermediate-period CCs are more likely to exhibit a significant relative photometric contribution from the companions, particularly at short wavelengths.
Companions of CCs are often hot main sequence (MS) stars, therefore making the CCs appear bluer.
This consequently biases the estimate of their color excesses and reddenings.
If not taken into account, their orbital displacement affects the trigonometric parallax measurements.

The census of the companions of CCs and RRLs is incomplete, due to the high contrast between the bright pulsators and their companions.
Stellar population synthesis models by \citetads{2015A&A...574A...2N} predict that the binary fraction of CCs is likely lower than for their MS progenitor, and that about half of the CCs are products of binary interactions.
This would be caused by interactions between the close-in companions and the Cepheid progenitors while they evolve on the red giant branch.
\citetads{2012Sci...337..444S} claims that 70\% to 100\% of O stars have companions, whereas \citetads{2015A&A...574A...2N} predict that only 35\% of Cepheids do. The binary fraction of Galactic CCs is thus a key observable to test the intermediate-mass  star formation and evolution scenarios.

Most of the detectable companions of CCs are hot dwarf stars, and their ultraviolet emission can dominate that of the cooler pulsator \citepads{2005AJ....130..789E, 2011AJ....142...87E}. Several CCs are in triple or quadruple systems,  e.g., \object{W Sgr} \citepads{2009AJ....137.3700E}, Polaris \citepads{2002ApJ...567.1121E, 2018ApJ...853...55B}, and Y Car \citepads{2005AJ....130..789E}.
Their spectroscopic signatures can also be observed, for example using the calcium-line method \citepads{2015MNRAS.448.3567K}.
A number of CC companions have been resolved using classical imaging \citepads{1994AJ....108..653E}, optical interferometry \citepads{2013A&A...552A..21G, 2014A&A...561L...3G}, adaptive optics \citepads{2014A&A...567A..60G}, or HST imaging \citepads{2008AJ....136.1137E, 2018ApJ...863..187E, 2016AJ....151..129E} paving the way to the measurement of their orbital parallaxes \citepads{Gallenne2018}.
A database of the known binary and multiple Galactic CCs is maintained at Konkoly Observatory\footnote{\url{http://www.konkoly.hu/CEP/intro.html}} \citepads{2003IBVS.5394....1S}.
Low-mass companions with surface magnetic fields generated by convection have been detected using their X-ray emission \citepads{2010AJ....139.1968E}, including for Cepheids in clusters \citepads{2014ApJ...785L..25E}.
Binary Cepheids have also been identified in the Magellanic Clouds \citepads{2012MNRAS.426.3148S} in particular in eclipsing binary systems \citepads{2002ApJ...573..338A, 2004ApJ...611.1100L, 2010Natur.468..542P, 2013MNRAS.436..953P, 2015ApJ...806...29P, 2015ApJ...815...28G} that provide extremely accurate stellar parameters \citepads{2018arXiv180601391P}.

The goal of the present work is to test for the presence of close-in companions of Galactic CCs and RRLs using the Gaia Second Data Release (hereafter GDR2; \citeads{2016A&A...595A...1G,2018A&A...616A...1G}).
Our CC and RRL samples are presented in Sect.~\ref{samples}.
In Sect.~\ref{pmbin} we use the GDR2 position and proper motion (PM) measurements together with the Hipparcos catalog \citepads{1997A&A...323L..49P, 2007ASSL..350.....V} to search for PM anomalies.
In the companion Paper~II \citepads{Kervella2018b} we search the GDR2 for common PM stars located near the CCs and RRLs, and we test the possibility that they are gravitationally bound. We postpone the discussion of individual stars to Paper~II.

\section{Selected samples\label{samples}}

We present in this section the sample of CCs and RRLs  selected for  our present PM analysis (Paper~I) and for the search for resolved common proper motion companions presented in Paper~II.
In particular, we  detail our choice of parallax values and the systematic corrections that we applied to the different data sets (Sect.~\ref{gdr2corrections}).

\subsection{Cepheids}

We chose the sample of 455 Galactic CCs assembled by \citetads{2000A&AS..143..211B}.
We uniformly adopted the CC parallaxes from the GDR2, which we corrected following the procedure detailed in Sect.~\ref{gdr2corrections}, except for four stars (U\,Aql, R\,Cru, SU\,Cru, and Y\,Sgr) for which we adopted the Hipparcos parallax from \citetads{2007ASSL..350.....V}.
For $\delta$\,Cep, we adopted the GDR2 parallax of its physical companion $\delta$\,Cep~B, as detailed in Paper~II.
For RY\,Vel, whose GDR2 and Hipparcos parallaxes are negative, we adopted the photometric distance of \citetads{2000A&AS..143..211B} based on multicolor period--luminosity relations, renormalized to the LMC distance modulus established by \citetads{2013Natur.495...76P}, giving $\varpi = 0.39 \pm 0.06$\,milliarcseconds (mas).
We add to the sample the short-period double-mode pulsator \object{Y Car}, which is a known triple system \citepads{2005AJ....130..789E}. We adopt the distance modulus of $\mu = 10.8 \pm 0.3$ determined by \citetads{1992ApJ...385..680E}, corresponding to a parallax of $\varpi = 0.69 \pm 0.10$\,mas, i.e., with a $\pm 15\%$ uncertainty. Although the membership of \object{Y Car} to the open cluster \object{ASCC 60} is listed as inconclusive by \citetads{2013MNRAS.434.2238A}, the distance of the cluster (1.1\,kpc) determined by \citetads{2016A&A...585A.101K} does not exclude this possibility. The GDR2 parallax value ($\varpi = 0.301 \pm 0.035$\,mas) is likely unreliable, possibly due to the astrometric wobble of the center of light of the system.

Out of the 455 CCs present in the \citetads{2000A&AS..143..211B} catalog plus \object{Y Car}, {\NtotalpmanomalyCEP} are present in the Hipparcos catalog and were tested for the presence of a proper motion anomaly (hereafter PMa). The remaining stars are usually fainter than the Hipparcos magnitude limit.

\subsection{RR Lyrae}

We extracted the RR Lyrae type variables from the General Catalogue of Variable Stars \citepads{2017ARep...61...80S}, which comprises 8509 stars.
Only 198 of these stars are present in the Hipparcos catalog and therefore suitable for the search for companions from their PMa.
We adopt the GDR2 parallaxes of RRLs, uniformly corrected following the procedure detailed in Sect.~\ref{gdr2corrections}.
For \object{RR Lyr} itself, which is absent from the GDR2 catalog, we adopt the TGAS parallax from the GDR1 of $\varpi[\mathrm{RR\,Lyr}] = 3.64 \pm 0.23$\,mas \citepads{2014A&A...571A..85M, 2016A&A...595A...1G}.

We processed all the stars present in the selected catalogs (456 CCs and 198 RRLs), but the variability class of some targets is incorrect, and we present the results related to these objects separately from CCs and RRLs.

\subsection{Gaia DR2 basic corrections and quality control\label{gdr2corrections}}

The GDR2 parallaxes are affected by a mean global zero point (ZP) offset \citepads{Lindegren18}.
Examples of determinations of the GDR2 ZP include for instance the work by \citetads{2018ApJ...861..126R}, who derived a value of $-46 \pm 13\,\mu$as specifically for CCs, and \citetads{2018MNRAS.481.1195M} who obtained $-56 \pm 6\,\mu$as for RRLs.
\citetads{2018A&A...616A..17A} list a statistically identical ZP value to that of \citetads{2018MNRAS.481.1195M} for the full sample of GDR2 RRLs ($-56 \pm 5\,\mu$as; their Table~1).
However, their ZP for the restricted sample of RRLs present in the General Catalogue of Variable Stars (GCVS; \citeads{2009yCat....102025S}) is $-33 \pm 9\,\mu$as.
For CCs, \citetads{2018A&A...616A..17A} obtain a ZP offset of $-32\,\mu$as.

The choice of ZP does not affect the PM anomaly and consequently the detected binaries in the present paper.
It has however an influence on the masses of the companions and their linear orbital radii, which are inversely proportional to the parallax.
The determination of the ZP of GDR2 is a complex question that is beyond the scope of the present work.
We therefore systematically corrected the GDR2 parallaxes of CCs and RRLs by adding a constant $\Delta \varpi_\mathrm{G2} = +29\,\mu$as offset to the catalog values, as recommended by \citetads{Lindegren18} and \citetads{2018arXiv180409376L}. This value corresponds to a sky average derived from quasar measurements, and is compatible with the $\Delta \varpi_\mathrm{G2}$ obtained by \citetads{2018A&A...616A..17A}  for CCs and for RRLs.
A future revision of the GDR2 ZP to a new value $\Delta \varpi_\mathrm{G2}^*$ can be used to correct the determined companion masses $m_2$ to new values $m_2^*$ through the simple multiplication
\begin{equation}
m_2^* = m_2 \ \frac{\varpi_\mathrm{G2} + \Delta \varpi_\mathrm{G2}}{\varpi_\mathrm{G2} + \Delta \varpi_\mathrm{G2}^*}
,\end{equation}
where $\varpi_\mathrm{G2}$ is the uncorrected parallax from the GDR2 catalog.
We note that choosing an offset correction of $\Delta \varpi_\mathrm{G2}=+56\,\mu$as instead of $+29\,\mu$as has a negligible impact on all the CC and RRL companion masses (Sect.~\ref{CEP-PMa} and \ref{rrlyr-results}) within their error bars.
We implemented the correction of the parallax uncertainties described in Eq. A.6 of \citetads{Lindegren18}, as recommended by \citetads{2018A&A...616A..17A}.

We corrected the GDR2 PM vectors for the rotation of the Gaia reference frame \citepads{Mignard18} reported by \citetads{Lindegren18} using the  expressions
\begin{gather}
\mu_{\alpha, \mathrm{corr}} = \mu_{\alpha} + w_x \sin(\delta) \cos(\alpha) + w_y \sin(\delta) \sin(\alpha) - w_z \cos(\delta) ,\\
\mu_{\delta, \mathrm{corr}} = \mu_{\delta} - w_x \sin(\alpha) + w_y \cos(\alpha),
\end{gather}
where $w_x = -0.086 \pm 0.025$\,mas\,a$^{-1}$, $w_y = -0.114 \pm 0.025$\,mas\,a$^{-1}$, and $w_z = -0.037 \pm 0.025$\,mas\,a$^{-1}$.
As discussed by \citetads{Lindegren18} the systematic uncertainty on the GDR2 PM vectors is limited to $\sigma_\mathrm{sys}(\mu) = 66\,\mu$as\,a$^{-1}$ per component for small separations (see also \citetads{2018A&A...616A..17A} and \citeads{2018arXiv180409376L}). This is possibly lower in reality, but we conservatively added quadratically this systematic uncertainty to the stated PM error bars of both RA and Dec axes.

Although this is not a requirement for the present Paper~I, we corrected the $G$-band magnitudes using the expression $G_\mathrm{corr} = 0.0505 + 0.9966\,G$ from \citetads{2018MNRAS.479L.102C} in view of the calibration of the resolved companion magnitudes in Paper~II.
The validity of this correction is demonstrated over the range $6 \lesssim G \lesssim 16.5$, but we also apply it to fainter stars.
The amplitude of the correction is at most 30\,mmag for a $G=6$ star, and therefore of marginal importance for our purpose.
The photometry of the stars with $G<6$ is unreliable due to saturation, but we did not find such bright candidate companions.

We tested the GDR2 record of the stars of our samples following the three quality criteria defined by \citetads{2018A&A...616A..17A} (their Sect. 4.1): 
 (1) a reduced $\chi^2$ of the Gaia astrometric fit below a limit dependent on the $G$ magnitude (e.g., $\chi^2_\mathrm{red}<8$ for a $G=10$ magnitude star), (2) a photometric $G_{BP}-G_{RP}$ flux excess factor within acceptable color-dependent limits, and (3) more than six visibility periods.
The stars that do not satisfy these three criteria are flagged with a $\dag$ symbol in Table~\ref{pm_binaries1}.
We also computed the reduced unit weight equivalent noise (RUWE\footnote{\url{https://www.cosmos.esa.int/web/gaia/dr2-known-issues}}; denoted $\varrho$ in the following, see also \citeads{2018arXiv181108902K}) of the GDR2 astrometric solution. 
The RUWE is a combination of the astrometric $\chi^2$, the number of good observations $N$, the $G$ magnitude, and the color index $C = G_\mathrm{BP} - G_\mathrm{RP}$.
In Table~\ref{pm_binaries1} and the  following the stars for which $\varrho > 1.4$ (as recommended by Lindegren) are flagged with a $\ddag$ symbol.

Binary stars present a natural discrepancy in the $\chi^2$ of their astrometric model fit, due to the present assumption in the GDR2 astrometric model that all stars are single. As a consequence, they are more likely to not fulfill  quality criterion (1) of \citetads{2018A&A...616A..17A} and exhibit $\varrho>1.4$. These quality indicators can thus be viewed as {de facto} indicators, however imperfect,  of astrometric binarity. To prevent the rejection of actual binary stars, we therefore kept all the stars in our analysis, including those with quality flags (we provide the flag information). Future Gaia data releases will include the binarity in the astrometric fit, and therefore provide a separate view of the contributions of binarity and instrument noise to the $\chi^2$.

\section{Binarity from proper motion anomaly\label{pmbin}}

\subsection{Proper motion anomaly\label{pmanomaly}}

\begin{figure}
\centering
\includegraphics[width=7cm]{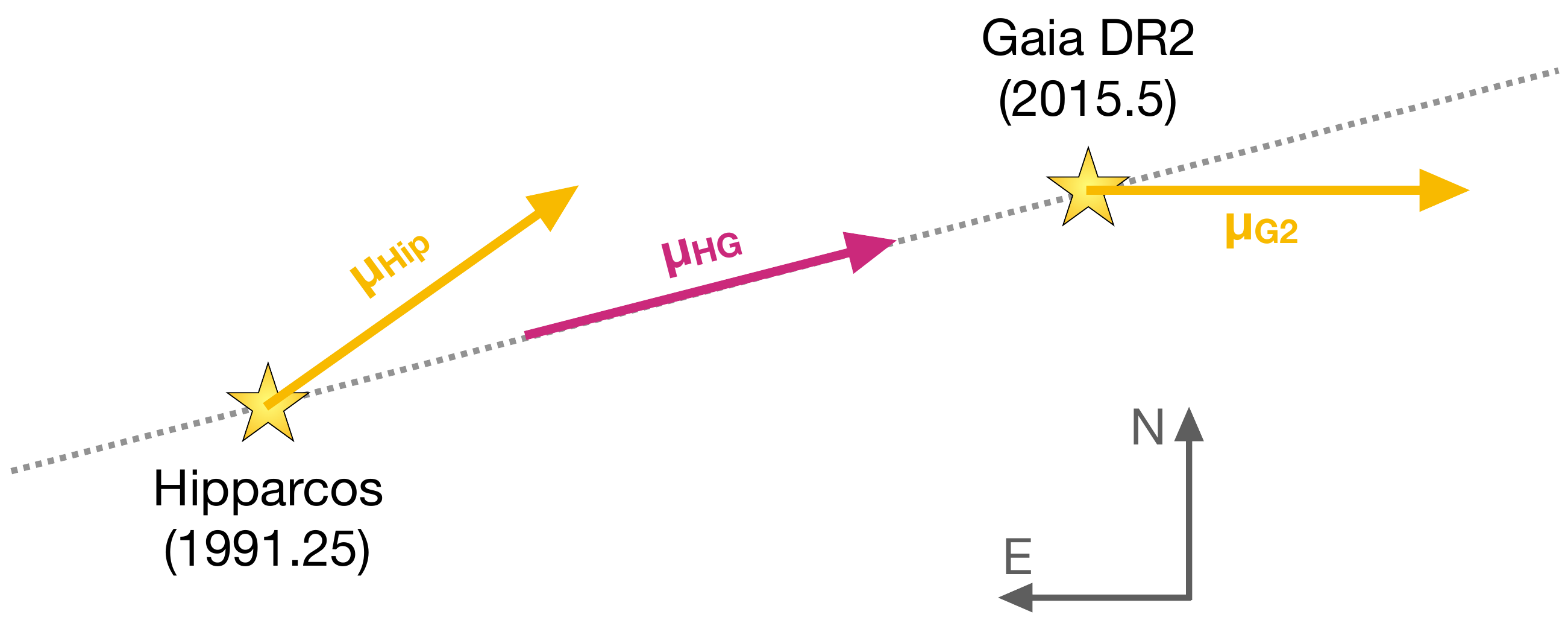}
\caption{Principle of the search for a proper motion anomaly. $\mu_\mathrm{Hip}$ designates the Hipparcos proper motion vector (epoch 1991.25), $\mu_\mathrm{HG}$ the mean proper motion vector between the Hipparcos and Gaia DR2 positions, and $\mu_\mathrm{G2}$ is the Gaia DR2 proper motion vector (epoch 2015.5).\label{PM-anomaly}}
\end{figure}

The principle of our search for close-in orbiting companions is to look for a difference in PM vector between the mean PM computed from the Hipparcos (1991.25) and GDR2 (2015.5) astrometric $(\alpha,\delta)$ positions on the one hand (hereafter $\mu_\mathrm{HG}$, for Hipparcos-Gaia) and the individual PM vectors $\mu_\mathrm{Hip}$ and $\mu_\mathrm{G2}$ respectively from the Hipparcos and GDR2 catalogs on the other hand.
This approach to compare the long-term to short-term PM vectors has historically been employed by \citetads{1844MNRAS...6R.136B} to discover the white dwarf companion of Sirius. It was also applied recently to various types of stars by \citetads{1999A&A...346..675W}, \citetads{2004ASPC..318..141J}, \citetads{2007A&A...464..377F}, \citetads{2008ApJ...687..566M}, \citetads{2018ApJS..239...31B},  \citetads{2018arXiv181107285B}, \citetads{2018arXiv181108902K}, and \citetads{2018NatAs...2..883S}.
Figure~\ref{PM-anomaly} shows the definition of the different PM vectors considered in the present work.
The combination of Gaia and Hipparcos data has already been used after Gaia DR1 \citepads{2016A&A...595A...4L} to produce the TGAS catalog \citepads{2014A&A...571A..85M, 2015A&A...574A.115M}. A description of the sources of uncertainty to take into account in the combination of these two catalogs is presented by \citetads{2018IAUS..330...41L}.

We identify $\mu_\mathrm{HG}$ to the projected velocity vector of the center of mass, while $\mu_\mathrm{Hip}$ and $\mu_\mathrm{G2}$ represent the projected velocity vector of the photocenter of the system at the Hipparcos and GDR2 epochs, respectively.
For single stars that have a linear uniform space motion, these three projected vectors have the same constant direction and norm (neglecting the variable spherical projection effects for such distant stars).
The presence of an orbiting companion will displace the photocenter away from the center of mass, due to the difference between the mass ratio and the flux ratio of the two stars.
In this case, the photocenter will revolve around their center of mass following a ``virtual orbit'' with a semimajor axis $a^\prime$
\begin{equation}
a^\prime = \frac{a\, L_1}{L_1+L_2}
,\end{equation}
where $a$ is the semimajor axis of the physical orbit of the primary star around the center of mass, $L_1$ its flux, and $L_2$ the flux of the secondary component.
As the Hipparcos and Gaia missions measure the PM of the photocenter, a deviation will appear with the PM of the center of mass.

The photocenter of a binary system comprising a CC is usually very close to the CC due to the high brightness of supergiants compared to their companions, that are usually MS dwarfs ($L_2 \ll L_1$).
For RRLs, the flux of the companion stars is also small; although the RRLs are less luminous than CCs, their companions are also significantly fainter (compact objects, very low-mass dwarfs) than their CC counterparts as they are very old stellar systems.
In the following, we  uniformly assume that the photocenter of the system is coincident with the position of the CC or RRL.
This assumption results in a systematic underestimation of the true tangential orbital velocity of the Cepheid by a factor $L_1 / (L_1+L_2)$.

We define the signal-to-noise ratio of the PMa of the Hipparcos/GDR2 measurements with respect to the mean PM $\mu_\mathrm{HG}$ as
\begin{equation}
\Delta_\mathrm{Hip/G2} = \frac{\mu_\mathrm{Hip/G2} - \mu_\mathrm{HG}}{\sqrt{\sigma_\mathrm{\mu\,Hip/G2}^2 + \sigma_\mathrm{\mu\,HG}^2 - C}},
\end{equation}
where $C = 2\,\rho\,\sigma_\mathrm{\mu\,Hip/G2}\,\sigma_\mathrm{\mu\,HG}$
corresponds to the correlation term (with a degree of correlation $\rho$) between $\mu_\mathrm{HG}$ and the PM vectors from the Hip/GDR2 catalogs.
These two quantities are correlated as the astrometric positions $(\alpha,\delta)$ in the Hip/GDR2 catalogs, which are used to compute $\mu_\mathrm{HG}$, and are themselves correlated to the $\mu_\mathrm{Hip/G2}$ PM vector coordinates.

However, since the position uncertainty intervenes in its computation with a divisive factor 24.25 (difference in years between the Hipparcos and GDR2 epochs), $C$ is much smaller than the $\mu_\mathrm{Hip}^2$ variance in $\Delta_\mathrm{Hip}$ and than the Hipparcos positional variance in $\Delta_\mathrm{G2}$, and can thus be neglected in both cases. 

For the identification of candidate binary stars, we consider the maximum of the two values  $\Delta = \max (\Delta_\mathrm{Hip}, \Delta_\mathrm{G2})$. In general, for a given star showing a PMa, the signal-to-noise ratio $\Delta_\mathrm{G2}$ is significantly higher than $\Delta_\mathrm{Hip}$ thanks to the higher accuracy of Gaia (typically by one order of magnitude). However,  depending on the configuration of the orbit and the orbital phase, a PMa may be detectable at the Hipparcos epoch and not at the GDR2 epoch.

\subsection{Constraints on companion properties\label{companion-properties}}

\subsubsection{Levels of analysis\label{levels}}

Several  levels of analysis can be achieved, depending on the available observational constraints:
\begin{enumerate}
\item \textit{Proper motion anomaly only:}
Knowing the parallax, $\Delta \mu$ gives the 2D tangential linear velocity $\vec{v_\mathrm{tan}}$ of the target in the center-of-mass referential at the measurement epoch.
This is a projection of the true 3D velocity vector of the star on the plane of the sky (i.e., the plane perpendicular to the line of sight containing the star), and therefore its norm is a lower limit of its orbital speed.
With an a priori estimate of the mass of the target (Sect.~\ref{apriorimasses}) and an additional hypothesis on the mass ratio $q$ of the binary (Sect.~\ref{apriorimassratio}), we derive a range of maximum semimajor axes and orbital periods using the expressions
\begin{equation}
a_\mathrm{max} = \frac{G\,m_1\,(1+q)}{v_\mathrm{tan}^2}, \ \ \ P_\mathrm{max} = \frac{2\pi\,a}{v_\mathrm{tan}}
\end{equation}
Here we implicitly assume that the orbit is circular.
For this simplified analysis, we considered only the PMa vectors determined from the GDR2. The PMa vectors from the Hipparcos catalog generally have one order of magnitude lower accuracy than those computed from the GDR2, and therefore provide limited constraints on the orbital radii and orbital periods of the companions.

\medskip
\item \textit{Proper motion anomaly and radial velocity:}
When the parameters of the spectroscopic orbit ($P, e, \omega, K$) are known, the knowledge of the orbital radial velocity $v_\mathrm{r}$ of the target at the same epoch as the PMa vector, together with the parallax, gives its complete 3D velocity vector $\vec{v} = [v_{\alpha},v_{\delta},v_\mathrm{r}]$.
The availability of two orbital velocity vectors, $\vec{v_\mathrm{Hip}}$ and $\vec{v_\mathrm{G2}}$, gives access to the inclination $i$ of the orbital plane and the longitude of the ascending node $\Omega$ through a cross product $\vec{v_\bot} = [v_{\bot \alpha},v_{\bot \delta},v_{\bot \mathrm{r}}] = \vec{v_\mathrm{Hip}} \times \vec{v_\mathrm{G2}}$:
\begin{equation}
i = \arccos \left( \frac{v_{\bot \mathrm{r}}}{|\vec{v_\bot}|} \right), \ \ \ \Omega = \arctan \left( \frac{v_{\bot \alpha}}{v_{\bot \delta}} \right) - 90^\circ.
\end{equation}
This resolves the $\sin(i)$ degeneracy, thus allowing us to derive the full set of orbital parameters.
With a prediction of the primary mass (Sect.~\ref{apriorimasses}), the mass $m_2$ of the secondary star can then be determined.

\end{enumerate}

We refer in the following to the level of analysis that we can achieve on a given target using, e.g., ``level 2'' to designate the systems with two PMa vectors and the corresponding radial velocities.

The Hipparcos observations were conducted between 7 November 1989 and 18 March 1993 \citepads{1997A&A...323L..49P}, i.e., covering 1227\,d.
The GDR2 catalog values are based on data collected between 25 July 2014 and 23 May 2016 \citepads{2018A&A...616A...1G}, i.e., covering 668\,d.
This means that the PM vectors from these two catalogs are not instantaneous, but represent a weighted average over these observing windows that depends on the distribution of the individual observed transits.
For binary systems with orbital periods shorter than these observing windows, the measured PMa is still a valuable tracer of binarity,
but due to the integration of more than one orbital cycle, the PMa vector coordinates are smoothed by the observing window.
As a consequence, the determination of the orbital parameters together with the spectroscopic orbit may be biased (see, e.g., the case of S\,Mus discussed in Sect.~\ref{pmvalidation}).
For short-period companions, the error bars of the Hipparcos and GDR2 PM vectors incorporate the residual wobble due to the several orbital cycles covered during the observing window \citepads{2018arXiv181108902K}.
The smoothing of the PMa signal also results in a significant decrease in the sensitivity of this indicator to orbiting companions for orbital periods $\lesssim 1000$\,days. The inclusion of binary fitting in the astrometric solution of future Gaia data releases will allow  this limitation to be waived (see, e.g., the recent work by \citeads{2018NatAs...2..883S} using Hipparcos epoch astrometry).

Cepheid binary systems with fully determined orbital parameters from the combination of visual (from classical imaging or optical interferometry) and spectroscopic orbit are still rare.
To date only \object{V1334 Cyg} has a fully determined, high-precision set of orbital parameters \citepads{Gallenne2018} including the masses of both components to 3\% accuracy and their distance to 1\% accuracy.
This favorable configuration provides a stringent test of the reliability of the PMa analysis.
In addition, \citetads{2018arXiv181209989G} recently obtained high-accuracy interferometric astrometric orbits from interferometry for U\,Aql and S\,Mus. We briefly discuss them in Sect.~\ref{pmvalidation} together with V1334 Cyg.

\subsubsection{A priori mass estimates\label{apriorimasses}}

The masses of the CCs were approximated using a combination of the theoretical period-luminosity-radius relation for fundamental mode pulsators by \citetads{2005ApJ...629.1021C} and the period--radius relation calibrated by \citetads{2017A&A...608A..18G}.
We did not ``fundamentalize'' the periods of the first overtone pulsators.
This is a very simple approach, but it provides sufficiently accurate estimates of the Cepheid masses for our purpose (companion mass and escape velocity estimates, see Paper~II).
For the short-period, first overtone pulsator \object{V1334 Cyg} ($P=3.33$\,d), the agreement between the prediction ($4.6\,M_\odot$) and the determined mass by G18 ($4.29\,M_\odot$) is satisfactory (8\%).
For the fundamental mode long-period pulsator $\ell$\,Car, the agreement is good between the predicted value ($8.4 M_\odot$), the range of $8-10\,M_\odot$ defined by \citetads{2016ApJ...824....1N} and the $9\,M_\odot$ estimate given by \citetads{2016MNRAS.455.4231A}.
For \object{Polaris}~Aa, which is a first overtone pulsator, the predicted mass is $m[\mathrm{Polaris\ Aa}] = 4.8\,M_\odot$, significantly lower than the $7\,M_\odot$ estimate by \citetads{2018A&A...611L...7A}. The estimate, however,  was derived assuming the HST/FGS parallax from \citetads{2018ApJ...853...55B}, which is underestimated \citepads{2515-5172-2-3-126}, and therefore the mass is likely overestimated.
From the astrometric monitoring of the orbit of \object{Polaris B}, \citetads{2018ApJ...863..187E} derive a value of $m[\mathrm{Polaris\ Aa}] = 3.45 \pm 0.75\,M_\odot$, which is lower than our estimate but compatible within the uncertainties.
In the following analysis, we adopt a conservative $\pm 15\%$ uncertainty on the predicted masses of the CCs of our sample.

For RRLs, we adopt a uniform mass of $0.6 \pm 0.1\,M_\odot$ ($\pm 15\%$) independent of the period. This conservatively covers the full range of possible masses predicted by the mass-metallicity relation of \citetads{1998A&A...333..571J}:
\begin{equation}
\log m = -0.328 - 0.062\,[\mathrm{Fe/H}]\ \ (\sigma = 0.019).
\end{equation}

\subsubsection{Mass ratio\label{apriorimassratio}}

Mass ratios of known multiple CCs are reviewed by \citetads{2015AJ....150...13E}. They are distributed mostly uniformly between 0 and 1. The CCs with determined companion masses  usually have mass ratios $q<1$ for MS companions, as the CC has to be more evolved than the companion. However, mass ratios larger than one are exceptionally possible when the companions are themselves binary systems,  for  example \object{AW Per} \citepads{2000AJ....120..407E, 2016Obs...136..209G}.

For RRLs, the only binary known with confidence is \object{TU UMa,} for which \citetads{2016A&A...589A..94L} estimate $m_1=0.55\,M_\odot$ and obtain a minimum mass for the companion of $m_2 = 0.34\,M_\odot$, which corresponds to a mass ratio $q = m_2/m_1 = 0.6$. We note however that the true mass of \object{TU UMa B} is significantly higher than this minimum value (Sect.~\ref{rrlyr-results}).

In absence of spectroscopic orbital parameters, we assume $q = 0.5 \pm 0.3$ for CC and RRL companions in the following discussion, with the pulsating star being the more massive. 

\subsection{Validation on V1334 Cyg \label{pmvalidation}}

\object{V1334 Cyg} is a short-period ($P=3.33$\,d) first overtone CC \citepads{2000AJ....119.3050E,2013A&A...552A..21G} that is a known spectroscopic and interferometric binary system \citepads{1995ApJ...445..393E,2000AJ....119.3050E,2013A&A...552A..21G}.
The full set of orbital parameters, masses and distance of \object{V1334 Cyg} have been determined with very high accuracy by G18.
This therefore provides us with an excellent test system (see Sect.~\ref{companion-properties}) to validate our approach based on PM anomalies.

\begin{figure}
\centering
\includegraphics[width=\hsize]{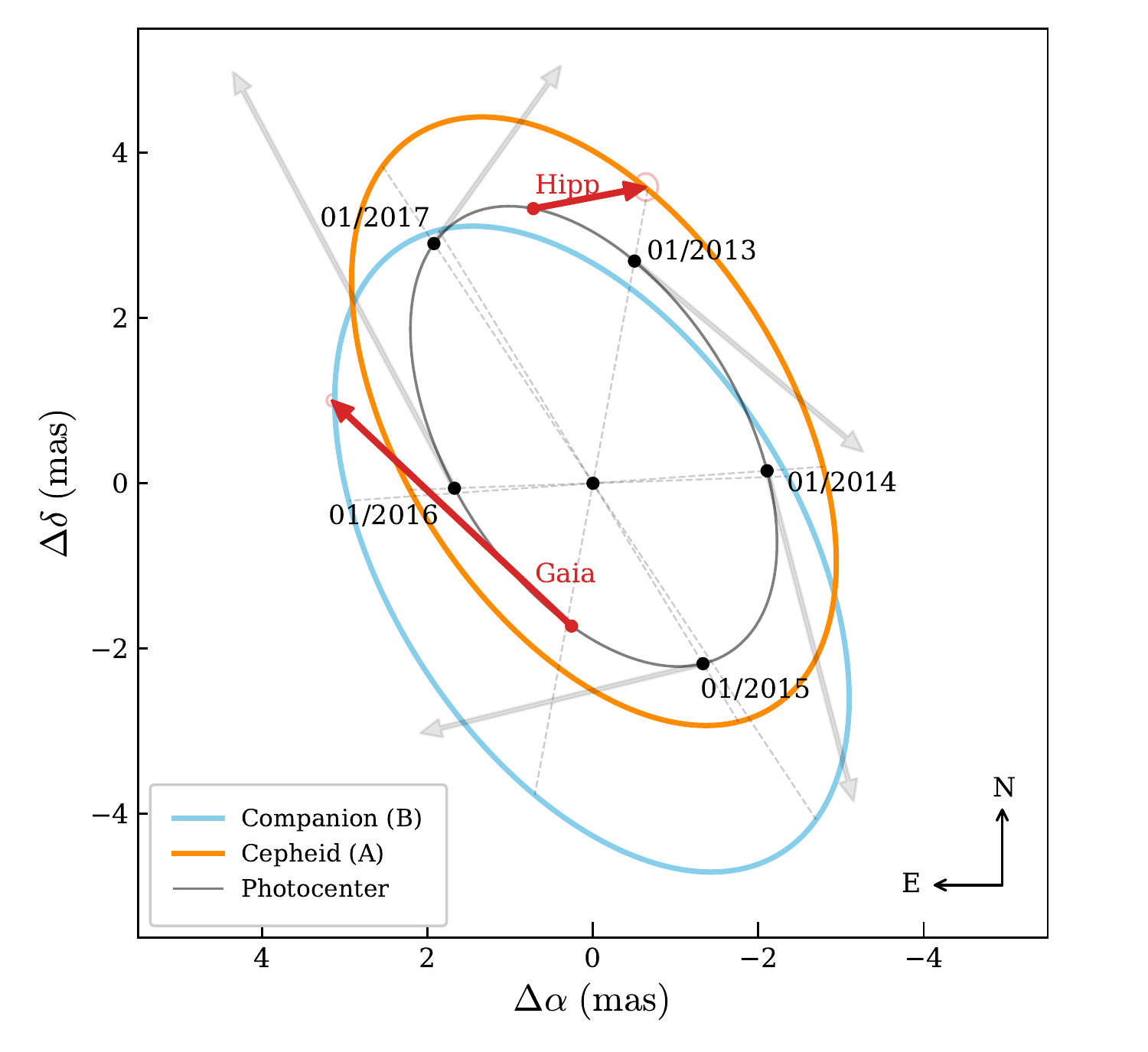}
\caption{Orbits of \object{V1334 Cyg} A (orange ellipse) and its companion B (light blue) around their common center of mass from \citetads{Gallenne2018}. The virtual orbit of the photocenter of the system is shown as a gray ellipse.
The measured tangential velocity vector (proper motion anomaly) is represented at the Hipparcos and Gaia epochs.\label{V1334Cyg-orbit}}
\end{figure}

Figure~\ref{V1334Cyg-orbit} shows the orbits of the two components around the barycenter from G18, as well as the Hip and GDR2 tangential velocity vectors (PM anomalies).
The agreement in position angle of the PM vectors with respect to the expected directions is satisfactory.
The GDR2 vector is consistent in terms of norm, but the Hip vector's norm is slower than expected.
To conduct a blind analysis we considered as input parameters only the spectroscopic orbital parameters determined by \citetads{2000AJ....119.3050E}.
The parameters that we derive are listed in Table~\ref{V1334Cyg-parameters}, together with the corresponding values found by G18 for comparison (in parentheses). The agreement is good on the inclination of the orbital plane $i$ ($1.2\sigma$) and the longitude of the ascending node $\Omega$ ($0.8\sigma$). The determined companion mass $m_2$ is also in good agreement ($0.3\sigma$).

The $i$ and $\Omega$ parameters are directly determined from the radial and PMa vectors at the Hip and GDR2 epochs. They are usually impossible to estimate without spatially resolving the system. This good consistency of the results between two fully independent approaches demonstrates the high potential of the PMa signal to determine orbital parameters from Gaia measurements.

It is interesting to note that we considered in this analysis that the photocenter is perfectly coincident with the Cepheid.  For \object{V1334 Cyg}, we know from \citetads{Gallenne2018} that its virtual orbit (Fig.~\ref{V1334Cyg-orbit}, gray ellipse) is $\approx 20$\% smaller than that of the CC (orange ellipse). Correcting {a posteriori} for this offset results in an increase of the companion mass to $m_2 = 3.9 \pm 0.6\,M_\odot$, within $0.2\sigma$ of the true mass of \object{V1334 Cyg B} ($4.0\,M_\odot$). The inclination $i$ is increased to $119 \pm 6^\circ$, which is also within $1\sigma$ of the value determined by \citetads{Gallenne2018}. This confirms that our hypothesis that the orbit of the photocenter is identical to that of the CC results in a systematic underestimation of the mass of the companions (Sect.~\ref{pmanomaly}).
However, \object{V1334 Cyg} is an extreme case as the relative brightness of \object{V1334 Cyg B} in the visible is not negligible ($\approx 10\%$). For most of the CC binaries considered here, the companions are much fainter, and the bias on the determined companion masses is negligible.

\begin{table}
 \caption{Parameters of the V1334 Cyg system from the combined analysis of the spectroscopic orbit of \citetads{2000AJ....119.3050E} and the proper motion anomaly vectors. The high accuracy values derived by \citetads{Gallenne2018} are given for each parameter in parentheses.}
 \label{V1334Cyg-parameters}
 \centering
 \renewcommand{\arraystretch}{1.2}
 \begin{tabular}{ll}
  \hline
  \hline
  \noalign{\smallskip}
\multicolumn{2}{l}{\textit{Adopted parameters}} \\
    Parallax  from GDR2 $\varpi$         &  $1.180_{\pm 0.066}$\,mas\ \ ($1.388_{\pm 0.015}$\,mas) \\
    Mass from P-M $m_1$     &  $4.6_{\pm 0.7}\,M_\odot$\ \ ($4.29_{\pm 0.13}\,M_\odot$) \\
  \hline
  \noalign{\smallskip}
\multicolumn{2}{l}{\textit{Parameters from \citetads{2000AJ....119.3050E}}} \\
    Orbital period  $P$                   &  $1937.5_{\pm 2.1}$\,d\ \ ($1932.8_{\pm 1.8}$\,d) \\
    Eccentricity  $e$                     &   $0.197_{\pm 0.009}$ \ \ ($0.233_{\pm 0.001}$) \\
    Arg. of periastron  $\omega$    &     $226.4_{\pm 2.9}$\,deg\ \ ($229.8_{\pm 0.3}\deg$) \\
    $v_\mathrm{r}$ amplitude   $K_1$ & $14.1_{0.1}$\,km\,s$^{-1}$ \ \ ($14.168_{0.014}$\,km\,s$^{-1}$) \\
    $v_\mathrm{r}$ at Hip epoch         &    $+9.86 \pm 0.41$\,km\,s$^{-1}$ \\
    $v_\mathrm{r}$ at GDR2 epoch        &    $-9.66 \pm 1.33$\,km\,s$^{-1}$ \\
  \hline
  \noalign{\smallskip}
\multicolumn{2}{l}{\textit{PMa vectors}} \\
    $\vec{\mu_\mathrm{Hip}}$  &   $[-1.36_{\pm 0.29}, +0.26_{\pm 0.33}]$\,mas\,a$^{-1}$ \\
    $\vec{\mu_\mathrm{G2}}$  &  $[+2.90_{\pm 0.12},  +2.73_{\pm 0.14}]$\,mas\,a$^{-1}$ \\
      \hline
  \noalign{\smallskip}
\multicolumn{2}{l}{\textit{Parameters from present analysis}} \\
    Inclination $i$   &  $118_{\pm 6}$\,deg\ \ $(124.94_{\pm 0.09}\deg)$ \\
    Semimajor axis  $a$                &    $6.18_{\pm 0.21}$\,au\ \ ($6.16_{\pm 0.07}$\,au) \\
    Ang. semimajor axis  $\theta$   &   $7.3_{\pm 0.5}$\,mas\ \ ($8.54_{\pm 0.04}$\,mas) \\
    Long. of asc. node  $\Omega$  &  $208_{\pm 6}$\,deg\ \ $(213.17_{\pm 0.35} \deg)$ \\
     Mass of secondary   $m_2$ &   $3.80_{\pm 0.57}\,M_\odot$\ \ $(4.04_{\pm 0.05}\,M_\odot)$ \\
  \hline
\end{tabular}
\end{table}

\citetads{2018arXiv181209989G} recently reported an orbital solution for U\,Aql and S\,Mus based on astrometric measurements obtained by interferometry.
They derived companion masses of $m_2 = 2.2 \pm 0.2\,M_\odot$ and $4.0 \pm 0.2\,M_\odot$, respectively for the two CCs.
For U\,Aql the agreement  with our estimate of $m_2 = 1.9 \pm 0.3\,M_\odot$ ($0.8\sigma$; Table~\ref{CEP-orbits-withRv}) is good.
For S\,Mus we obtain a companion mass of $m_2 = 2.2 \pm 0.3\,M_\odot$, significantly lower ($5\sigma$) than the \citetads{2018arXiv181209989G} value. This discrepancy arises from the orbital period of this system, which is significantly shorter ($P_\mathrm{orb} = 506$\,d) than the GDR2 and Hipparcos measurement windows.
As discussed in Sect.~\ref{levels}, this biases the corresponding PMa estimate, as testified by the classification of S\,Mus as ``preliminary'' in Table~\ref{CEP-orbits-withRv}.

\subsection{PM anomalies of Cepheids\label{CEP-PMa}}

\begin{figure}
\centering
\includegraphics[width=\hsize]{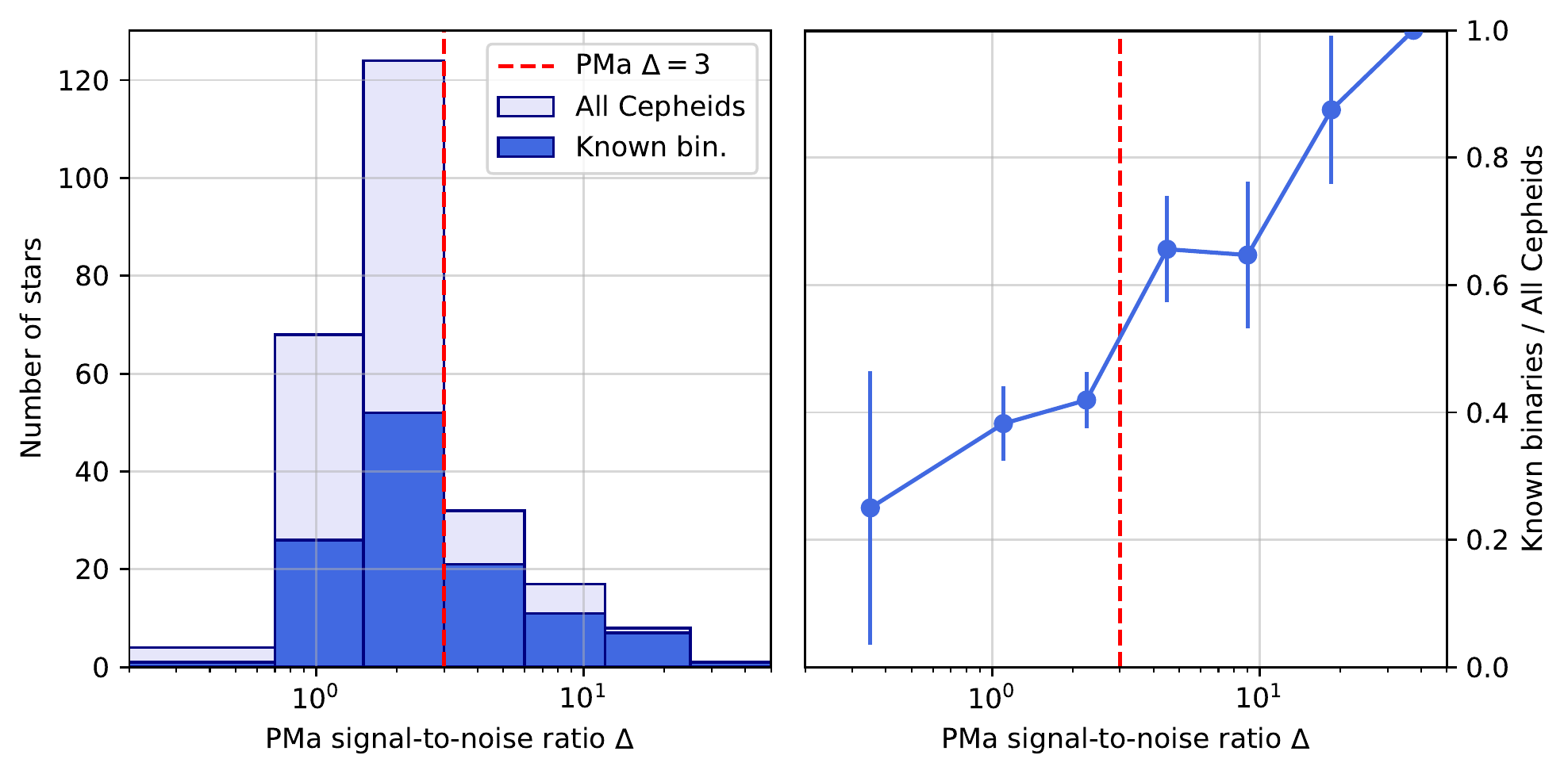}
\caption{Histogram of the detected PMa signal-to-noise ratios $\Delta$ of CCs (right panel) and fraction of known binaries with respect to the total number of Cepheids per bin (left panel). The dashed red lines mark our binary detection threshold of $\Delta=3$. \label{PMa-histo}}
\end{figure}

\begin{figure}
\centering
\includegraphics[width=\hsize]{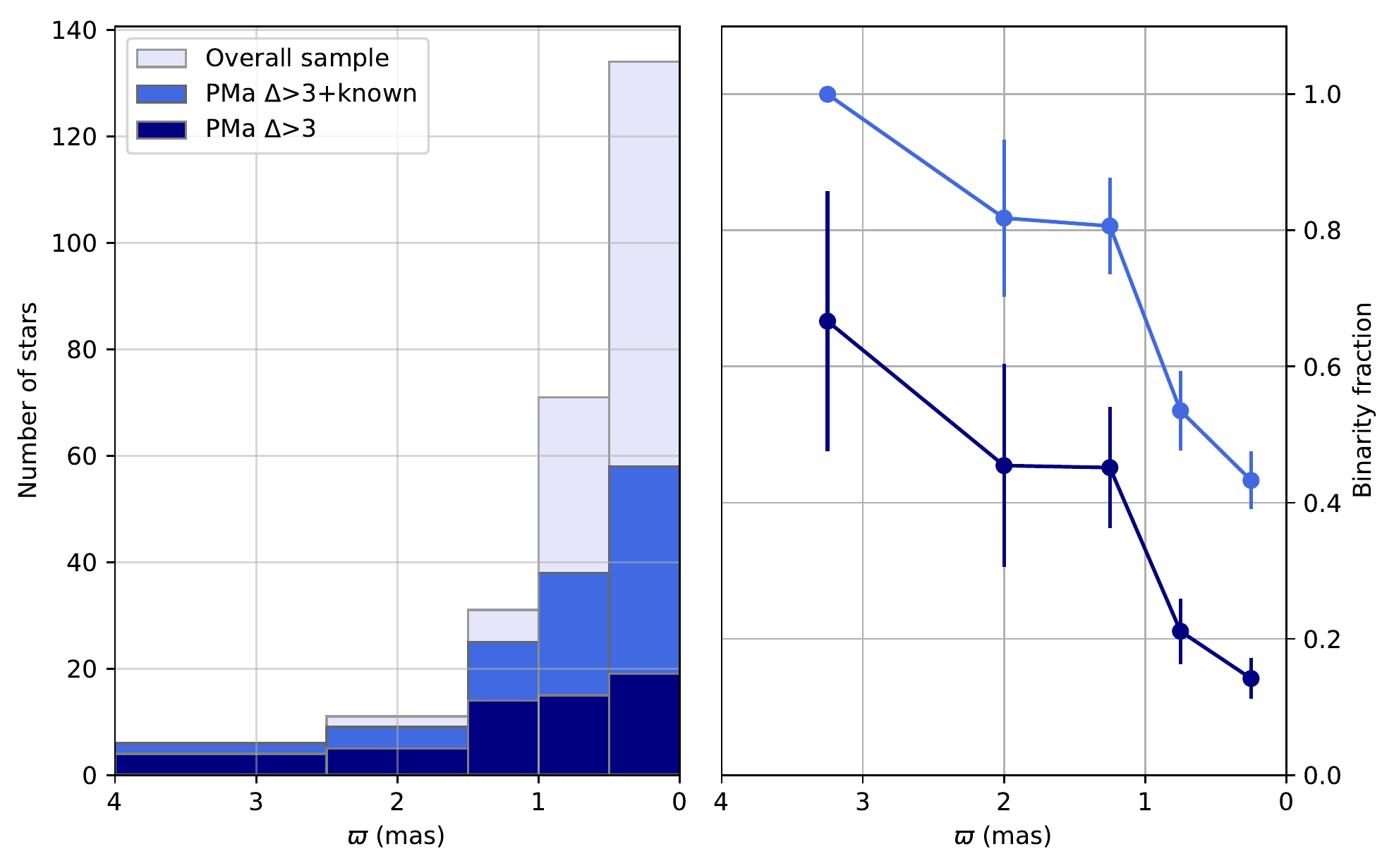}
\caption{\textit{Left:} Histogram of the Cepheids that show a proper motion anomaly ($\Delta>3$, dark blue), with the additional stars classified as binaries in the database maintained by \citetads{2003IBVS.5394....1S} (medium blue) and the overall sample (light blue) as a function of parallax.
\textit{Right:} Binary fraction as a function of parallax. The error bars represent the binomial proportion 68\% confidence interval.
\label{CEP-histo}}
\end{figure}

Tables \ref{pm_binaries1} to \ref{pm_binaries3} list the result of the search for PM anomalies on our selection of CCs.
We identify {\NhighpmanomalyCEP} stars with a high $\Delta>5$, {\NmidpmanomalyCEP} stars with $3<\Delta<5$, and {\NlowpmanomalyCEP} additional CCs with indications of a PM anomaly with $2<\Delta<3$.
The fraction of CCs in our sample of {\NtotalpmanomalyCEP} tested stars showing PMa at least at a $\Delta=3$ level is therefore {\pmanomalyfractionCEP}\%.
The histogram of the observed PMa signal-to-noise ratio $\Delta$ of the CC sample is presented in Fig.~\ref{PMa-histo} together with the fraction of known CCs for a given range of $\Delta$ values. We observe that among the CCs with $\Delta > 3$, approximately 70\% are previously known binary systems.
The histogram of the detected sources as a function of their parallax is presented in Fig.~\ref{CEP-histo}.
When combined with the known binary systems from the \citetads{2003IBVS.5394....1S} database the fraction of binary systems in CCs comes out at a high level.
Eight out of nine CCs in our sample within 500\,pc ($\varpi > 2$\,mas) are in binary systems: six exhibit $\Delta>3$ (U\,Aql; SU\,Cas; $\delta$\,Cep; SU\,Cru; $\beta$\,Dor; X\,Sgr) and two others are classified as binaries from \citetads{2003IBVS.5394....1S} ($\eta$\,Aql, $\Delta=1.72$; Y\,Sgr, $\Delta=2.98$).
Although $\zeta$\,Gem is also listed as a binary by \citetads{2003IBVS.5394....1S}, its visual companion is not physically related to the CC (Paper~II), and we therefore removed this star from our binary star count.
We note that SU\,Cas and $\delta$\,Cep are likely triple and quadruple systems, respectively, and Polaris, which is the nearest CC (but is not part of our sample), is a triple system (Paper~II).

Considering the 100 nearest CCs in our sample (with $\varpi > 0.56$\,mas), 32 stars show $\Delta>3$, and 31 others are known binaries from the \citetads{2003IBVS.5394....1S} database.
Four CCs in this sample (TV\,CMa, ER\,Car, V0532\,Cyg, and V0950\,Sco) are not classified as binaries by \citetads{2003IBVS.5394....1S} and have $\Delta<3$, but we report resolved gravitationally bound companions in Paper~II.
Altogether, we therefore obtain a minimum binary fraction of $P = 67\%$ for this sample of 100 nearby CCs.
Another way to estimate the binary fraction is to rely on an estimate of the completeness level $r$ of the PMa binary detections within our 100 CC sample. An approximation of $r$ is provided by the fraction of CCs with $\Delta > 3$ among the known binary CCs. We obtain a value of $r = 43\%$ for our sample that characterizes the mean efficiency of the PMa analysis to detect known CC binaries.
Applying this ratio to the detected PMa with $\Delta > 3$ gives an extrapolated binary fraction of $P = 31/0.43 = 72\%$, which is consistent with the overall minimum binarity previously determined.
The smoothly decreasing shape of the binary fraction curve shown in Fig.~\ref{CEP-histo} (right panel) is due to the  the sensitivity of the PMa technique in terms of companion mass being a linear function of the parallax.
This can be observed, for instance, by restricting our sample to the 50 closest CCs. This sample contains 24 stars with $\Delta>3$, and we derive $r = 49\%$, hence an extrapolated binary fraction of $P=98\%$ (in agreement with Fig.~\ref{CEP-histo} for nearby stars).

Considering the minimum $P$ values above, and taking into account the decreasing sensitivity of the PMa analysis with distance, we conclude that the actual binary fraction of CCs is probably above 80\%.
This fraction is consistent with the estimate by \citetads{2003IBVS.5394....1S} ($P \gtrsim 80\%$), but higher than observed by, among others,  \citetads{2016ApJS..226...18A} ($P = 32-52\%$) or \citetads{2012MNRAS.424.1925C} ($P \approx 40-70\%$ for CC progenitors) and predicted by \citetads{2015A&A...574A...2N} ($\approx 35\%$).

\begin{table*}
 \caption{Orbital parameters for Cepheid systems that have spectroscopic orbital parameters.
The lower part of the table lists the Cepheids with orbital periods shorter than 1000 days for which the orbital parameters and companion mass are poorly constrained (see Sect.~\ref{levels}).
A null value for the eccentricity $e$ and the argument of periastron $\omega$ indicates that the spectroscopic orbit was assumed to be circular.}
 \label{CEP-orbits-withRv}
 \small
 \centering
 \setlength{\tabcolsep}{5pt}
 \renewcommand{\arraystretch}{1.2}
 \begin{tabular}{lcllcrrrrrrcc}
\hline \hline
 Target & $\varpi$ & $P_\mathrm{orb}^\star$& MJD$_0^\star$& $e^\star$& $\omega^\star$& $K_1^\star$& $i$& $\Omega$& $a$& $a$& $m_1^\dag$& $m_2$\\
 & (mas)& (d) & && $(^\circ)$& (km\,s$^{-1}$) & ($^\circ$) &($^\circ$) & (au) & (mas) & ($M_\odot$) & ($M_\odot$) \\
\hline \noalign{\smallskip}
\object{U Aql} & $ 3.63_{ 0.96}$ & $ 1831_{ 6.5}$ & $ 47575_{ 8}$ & $0.193_{0.005}$ & $167_{ 2}$ & $ 8.4_{0.04}$ & $ 65_{ 10}$ & $345_{ 5}$ & $ 5.64_{ 0.22}$ & $ 20.47_{ 5.47}$ & $ 5.20_{ 0.78}$ & $ 1.94_{ 0.29}$ \\ 
\object{FF Aql} & $ 1.84_{ 0.11}$ & $ 1430_{ 2.6}$ & $ 58296_{ 14}$ & $0.061_{0.007}$ & $316_{ 4}$ & $ 4.8_{0.01}$ & $ 89_{ 7}$ & $ 61_{ 10}$ & $ 4.47_{ 0.19}$ & $ 8.22_{ 0.59}$ & $ 5.00_{ 0.75}$ & $ 0.83_{ 0.12}$ \\ 
\object{V0496 Aql} & $ 0.97_{ 0.05}$ & $ 1066_{ 1.9}$ & $ 45480_{ 17}$ & 0 & 0 & $ 3.6_{0.18}$ & $ 44_{ 7}$ & $283_{ 30}$ & $ 3.83_{ 0.17}$ & $ 3.72_{ 0.25}$ & $ 5.70_{ 0.85}$ & $ 0.89_{ 0.13}$ \\ 
\object{RX Cam} & $ 0.81_{ 0.04}$ & $ 1114_{ 0.5}$ & $ 45931_{ 2}$ & $0.459_{0.007}$ & $ 78_{ 1}$ & $14.3_{0.11}$ & $ 71_{ 6}$ & $ 91_{ 19}$ & $ 4.21_{ 0.15}$ & $ 3.41_{ 0.21}$ & $ 5.40_{ 0.81}$ & $ 2.61_{ 0.39}$ \\ 
\object{delta Cep} & $ 3.39_{ 0.05}$ & $ 2202_{ 6.3}$ & $ 55650_{ 25}$ & $0.674_{0.038}$ & $247_{ 5}$ & $ 1.5_{0.24}$ & $163_{ 14}$ & $ 83_{ 27}$ & $ 5.85_{ 0.24}$ & $ 19.86_{ 0.88}$ & $ 4.80_{ 0.72}$ & $ 0.72_{ 0.11}$ \\ 
\object{AX Cir} & $ 1.77_{ 0.34}$ & $ 6532_{ 25.0}$ & $ 48500_{ 60}$ & $0.190_{0.020}$ & $231_{ 8}$ & $10.0_{0.50}$ & $130_{ 6}$ & $ 3_{ 22}$ & $ 14.66_{ 0.53}$ & $ 26.00_{ 5.14}$ & $ 4.70_{ 0.70}$ & $ 5.15_{ 0.77}$ \\ 
\object{VZ Cyg} & $ 0.46_{ 0.03}$ & $ 2183_{ 10.0}$ & $ 44810_{ 36}$ & 0 & 0 & $ 3.0_{0.16}$ & $161_{ 9}$ & $356_{ 28}$ & $ 6.18_{ 0.22}$ & $ 2.87_{ 0.20}$ & $ 4.60_{ 0.69}$ & $ 2.00_{ 0.30}$ \\ 
\object{V1334 Cyg} & $ 1.18_{ 0.07}$ & $ 1938_{ 2.1}$ & $ 43606_{ 14}$ & $0.197_{0.009}$ & $226_{ 3}$ & $14.1_{0.10}$ & $118_{ 6}$ & $208_{ 6}$ & $ 6.18_{ 0.22}$ & $ 7.29_{ 0.49}$ & $ 4.60_{ 0.69}$ & $ 3.80_{ 0.57}$ \\ 
\object{T Mon} & $ 0.56_{ 0.07}$ & $ 32449_{ 726.0}$ & $ 49300_{ 143}$ & $0.414_{0.013}$ & $204_{ 3}$ & $ 8.4_{0.19}$ & $119_{ 17}$ & $ 87_{ 51}$ & $ 50.39_{ 1.90}$ & $ 28.32_{ 3.49}$ & $ 7.80_{ 1.17}$ & $ 8.41_{ 1.26}$ \\ 
\object{S Nor} & $ 1.09_{ 0.04}$ & $ 3584_{ 33.0}$ & $ 45638_{ 67}$ & 0 & 0 & $ 2.5_{0.35}$ & $154_{ 23}$ & $ 36_{ 43}$ & $ 8.87_{ 0.34}$ & $ 9.67_{ 0.53}$ & $ 5.70_{ 0.85}$ & $ 1.55_{ 0.23}$ \\ 
\object{AW Per} & $ 1.07_{ 0.06}$ & $ 13954_{ 181.0}$ & $ 52580_{ 28}$ & $0.474_{0.008}$ & $250_{ 1}$ & $10.3_{0.08}$ & $ 42_{ 6}$ & $ 56_{ 11}$ & $ 27.22_{ 1.01}$ & $ 29.15_{ 2.05}$ & $ 5.00_{ 0.75}$ & $ 8.82_{ 1.32}$ \\ 
\object{W Sgr} & $ 1.21_{ 0.41}$ & $ 1616_{ 11.0}$ & $ 57992_{ 20}$ & $0.197_{0.018}$ & $288_{ 6}$ & $ 1.6_{0.01}$ & $ 15_{ 12}$ & $ 59_{ 20}$ & $ 5.01_{ 0.21}$ & $ 6.05_{ 2.08}$ & $ 5.30_{ 0.79}$ & $ 1.12_{ 0.17}$ \\ 
\object{V0350 Sgr} & $ 1.01_{ 0.05}$ & $ 1472_{ 0.2}$ & $ 50526_{ 7}$ & $0.352_{0.001}$ & $284_{ 0}$ & $10.4_{0.01}$ & $ 35_{ 12}$ & $303_{ 38}$ & $ 5.16_{ 0.17}$ & $ 5.24_{ 0.30}$ & $ 4.70_{ 0.70}$ & $ 3.75_{ 0.56}$ \\ 
\object{V0636 Sco} & $ 1.01_{ 0.04}$ & $ 1323_{ 0.0}$ & $ 34411_{ 3}$ & $0.250_{0.001}$ & $290_{ 0}$ & $11.9_{0.01}$ & $ 94_{ 8}$ & $292_{ 14}$ & $ 4.59_{ 0.18}$ & $ 4.61_{ 0.27}$ & $ 5.10_{ 0.76}$ & $ 2.25_{ 0.34}$ \\ 
\object{U Vul} & $ 1.08_{ 0.04}$ & $ 2510_{ 2.8}$ & $ 44800_{ 16}$ & $0.675_{0.033}$ & $353_{ 4}$ & $ 3.6_{0.44}$ & $163_{ 7}$ & $320_{ 36}$ & $ 7.15_{ 0.25}$ & $ 7.74_{ 0.39}$ & $ 5.40_{ 0.81}$ & $ 2.35_{ 0.35}$ \\ 
 \noalign{\smallskip} \hline \noalign{\smallskip}
\object{Y Car} & $  0.69_{  0.10}$ & $   993_{   2.0}$ & $  45372_{     13}$ & $0.380_{0.020}$ & $129_{ 17}$ & $ 8.9_{0.30}$ & $150_{ 21}$  & $214_{ 38}$ & $  3.73_{  0.14}$ & $  2.57_{  0.40}$ & $  4.20_{  0.63}$ & $  2.82_{  0.42}$  \\ 
\object{YZ Car} & $ 0.35_{ 0.03}$ & $ 830_{ 0.2}$ & $ 53604_{ 5}$ & $0.027_{0.003}$ & $272_{ 2}$ & $10.2_{0.01}$ & $ 86_{ 13}$ & $ 60_{ 37}$ & $ 3.57_{ 0.13}$ & $ 1.25_{ 0.12}$ & $ 6.90_{ 1.03}$ & $ 1.93_{ 0.29}$ \\ 
\object{SU Cas} & $ 2.15_{ 0.08}$ & $ 407_{ 0.0}$ & $ 50278_{ 6}$ & 0 & 0 & $ 1.0_{0.08}$ & $ 48_{ 15}$ & $275_{ 17}$ & $ 1.65_{ 0.07}$ & $ 3.54_{ 0.21}$ & $ 3.50_{ 0.53}$ & $ 0.11_{ 0.02}$ \\ 
\object{BY Cas} & $ 0.51_{ 0.04}$ & $ 563_{ 5.0}$ & $ 49384_{ 5}$ & $0.220_{0.020}$ & $288_{ 10}$ & $ 9.1_{1.00}$ & $ 31_{ 22}$ & $188_{ 51}$ & $ 2.55_{ 0.09}$ & $ 1.30_{ 0.10}$ & $ 4.50_{ 0.67}$ & $ 2.44_{ 0.37}$ \\ 
\object{DL Cas} & $ 0.45_{ 0.03}$ & $ 684_{ 0.2}$ & $ 47161_{ 2}$ & $0.350_{0.006}$ & $ 27_{ 1}$ & $16.4_{0.11}$ & $141_{ 81}$ & $310_{ 95}$ & $ 3.28_{ 0.11}$ & $ 1.48_{ 0.12}$ & $ 5.40_{ 0.81}$ & $ 4.69_{ 0.70}$ \\ 
\object{XX Cen} & $ 0.57_{ 0.04}$ & $ 924_{ 1.1}$ & $ 44860_{ 8}$ & 0 & 0 & $ 4.5_{0.28}$ & $ 38_{ 13}$ & $164_{ 30}$ & $ 3.57_{ 0.14}$ & $ 2.02_{ 0.17}$ & $ 5.90_{ 0.89}$ & $ 1.23_{ 0.18}$ \\ 
\object{SU Cyg} & $ 1.20_{ 0.05}$ & $ 549_{ 0.0}$ & $ 43766_{ 1}$ & $0.350_{0.004}$ & $224_{ 1}$ & $29.8_{0.15}$ & $ 98_{ 7}$ & $100_{ 18}$ & $ 2.73_{ 0.09}$ & $ 3.27_{ 0.18}$ & $ 4.30_{ 0.64}$ & $ 4.70_{ 0.71}$ \\ 
\object{MW Cyg} & $ 0.73_{ 0.04}$ & $ 440_{ 0.2}$ & $ 48862_{ 15}$ & $0.140_{0.030}$ & $ 78_{ 13}$ & $ 6.4_{0.19}$ & $ 94_{ 20}$ & $237_{ 60}$ & $ 2.01_{ 0.08}$ & $ 1.46_{ 0.10}$ & $ 4.90_{ 0.73}$ & $ 0.72_{ 0.11}$ \\ 
\object{Z Lac} & $ 0.49_{ 0.04}$ & $ 383_{ 0.1}$ & $ 46582_{ 22}$ & $0.025_{0.012}$ & $344_{ 21}$ & $10.4_{0.10}$ & $ 95_{ 24}$ & $173_{ 80}$ & $ 2.00_{ 0.08}$ & $ 0.97_{ 0.09}$ & $ 5.90_{ 0.89}$ & $ 1.34_{ 0.20}$ \\ 
\object{S Mus} & $ 1.16_{ 0.12}$ & $ 506_{ 0.2}$ & $ 48556_{ 6}$ & $0.086_{0.004}$ & $194_{ 2}$ & $14.9_{0.01}$ & $ 80_{ 8}$ & $136_{ 70}$ & $ 2.48_{ 0.09}$ & $ 2.88_{ 0.31}$ & $ 5.70_{ 0.85}$ & $ 2.24_{ 0.34}$ \\ 
\object{S Sge} & $ 0.67_{ 0.09}$ & $ 676_{ 0.0}$ & $ 48010_{ 2}$ & $0.238_{0.005}$ & $203_{ 1}$ & $15.6_{0.06}$ & $ 60_{ 16}$ & $210_{ 9}$ & $ 3.08_{ 0.11}$ & $ 2.07_{ 0.30}$ & $ 5.50_{ 0.82}$ & $ 3.00_{ 0.45}$ \\ 
\object{X Sgr} & $ 3.46_{ 0.20}$ & $ 574_{ 0.6}$ & $ 48208_{ 19}$ & 0 & 0 & $ 2.3_{0.27}$ & $ 42_{ 18}$ & $236_{ 36}$ & $ 2.40_{ 0.12}$ & $ 8.32_{ 0.63}$ & $ 5.20_{ 0.78}$ & $ 0.43_{ 0.06}$ \\ 
\object{FN Vel} & $ 0.24_{ 0.04}$ & $ 472_{ 0.1}$ & $ 55936_{ 2}$ & $0.220_{0.010}$ & $266_{ 1}$ & $21.9_{0.08}$ & $ 93_{ 10}$ & $295_{ 22}$ & $ 2.36_{ 0.08}$ & $ 0.56_{ 0.09}$ & $ 4.80_{ 0.72}$ & $ 3.10_{ 0.47}$ \\ 
\noalign{\smallskip}
\hline
\end{tabular}
\tablefoot{$^\star$: spectroscopic orbital elements.}
\tablebib{The spectroscopic orbital elements were retrieved from the \citetads{2003IBVS.5394....1S} database, based on the following references:
\citetads{2013A&A...550A..70G} for V0496 Aql, VZ Cyg, MW Cyg, RX Cam, DL Cas; 
\citetads{2015ApJ...804..144A} for $\delta$ Cep;
\citetads{2008A&A...488...25G} for SU Cas, XX Cen, AX Cir, SU Cyg, Z Lac, T Mon, S Nor, S Sge, X Sgr, U Vul;
\citetads{2004MNRAS.350...95P} for Y Car;
\citetads{2016Obs...136..209G} for AW Per;
\citetads{Gallenne2018} for U Aql, FF Aql, V1334 Cyg, S Mus, W Sgr, V0350 Sgr, V0636 Sco;
\citetads{1995IBVS.4199....1G} for BY Cas;
\citetads{2013PhDT.......363A}\footnote{\url{https://archive-ouverte.unige.ch/unige:35356}} for FN Vel;
\citetads{2016ApJS..226...18A} for YZ Car.
}
\end{table*}

\begin{figure}
\centering
\includegraphics[width=\hsize]{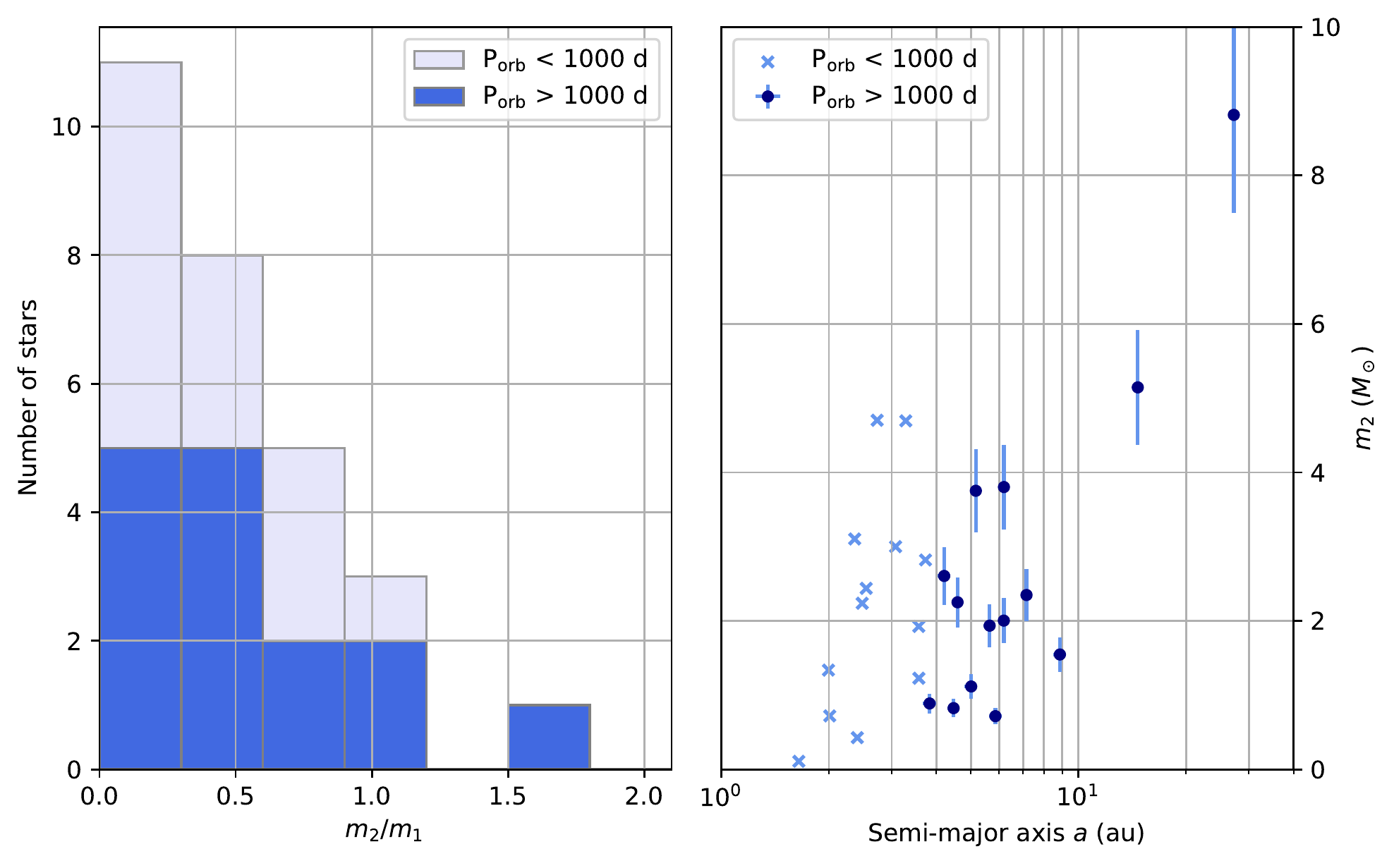}
\caption{\textit{Left:} Histogram of the mass ratios $q=m_2/m_1$ of the Cepheid systems with spectroscopic orbits. 
\textit{Right:} Distribution of the companion masses $m_2$ as a function of the orbital semimajor axis.
The ``preliminary'' systems in both panels correspond to the systems with uncertain parameters (lower part of Table~\ref{CEP-orbits-withRv}).
\label{CEP-histomassratio}}
\end{figure}

The orbital parameters and companion masses that we derive from the level 2 analysis of the systems that have spectroscopic orbits are presented in Table~\ref{CEP-orbits-withRv}.
The derived companion mass $m_2$ is inversely proportional to the adopted parallax $\varpi$ of the system. As a consequence, a revision of the Gaia parallaxes in the future data releases will result in a revision of the companion masses.
We note a good agreement of our values of $\Omega$ for \object{FF Aql} and \object{W Sgr} with the HST-FGS determinations by \citetads{2007AJ....133.1810B}, but a difference in the inclinations $i$, which may come from a different definition of this parameter.
\citetads{2015ApJ...804..144A} recently announced the discovery of a close-in companion of the prototype CC $\delta$\,Cep.
Adopting their spectroscopic orbital parameters, the mass $m_2 = 0.72 \pm 0.11\,M_\odot$ that we obtain for this companion is in good agreement with the range of $0.2-1.2\,M_\odot$ estimated by these authors.
We note, however, that due to its high brightness, the reliability of the GDR2 astrometric solution of $\delta$\,Cep is uncertain, and its PM vector may therefore be biased.
This companion mass should therefore be confirmed using a more accurate PM vector from a future Gaia data release.
We list in Table~\ref{CEP-orbits-withRv} the results for all targets, but the shorter periods ($\lesssim 1000$\,d) should be considered preliminary due to the PM vector smearing over the Hipparcos and GDR2 observing windows (Sect.~\ref{levels}).
A histogram of the mass ratios $q = m_2/m_1$ is presented in Fig.~\ref{CEP-histomassratio} (left panel).
We observe a high frequency of relatively  low-mass companions, with a median mass ratio $q = 0.4$.
This is consistent with the statistical estimates by \citetads{2017ApJS..230...15M} for CCs in binaries, which are based on the observational results by \citetads{2013AJ....146...93E} and \citetads{2015AJ....150...13E}.
The distribution of the masses as a function of the orbital semimajor axis is shown in Fig.~\ref{CEP-histomassratio} (right panel). The limited number of systems with a semimajor axis larger than 10\,au shows that the radial velocity technique has a higher sensitivity to short orbital periods. For the same companion mass, the astrometric detection technique is more sensitive to long periods and is therefore very complementary.

Selected properties of binary systems with CCs were compared with the synthetic population of 30 000 solar metallicity CCs in binaries, generated with the population synthesis code StarTrack \citepads{2002ApJ...572..407B,2008ApJS..174..223B}. The simulation confirms that the mass ratios (from less to more massive stars) form a uniform distribution over the entire Cepheid mass range ($5-10\,M_\odot$), with an average mass ratio of 0.5.
The distribution of the companion masses as a function of the orbital semimajor axis (Fig.~\ref{CEP-histomassratio}) is in good agreement with the population synthesis model; the model not only confirms that at  separations up to 7 au ($1500\,R_\odot$) the companions have masses smaller than $6\,M_\odot$, it also predicts more diverse companion masses (with the upper limit of $10\,M_\odot$) for larger separations.
Within the synthetic population, 99\% of all companions to CCs are MS stars, and only 1\% of the companions are giant stars. Among MS companions 75\% have spectral types B and A, which introduce a non-negligible photometric contribution, particularly at short wavelengths.
On average, the difference in magnitude between the binary system (CC with companion) and the CC alone is as large as 0.375~mag in the U-band, and decreases with  increasing wavelength: 0.127~mag ($B$), 0.053~mag ($V$), 0.037~mag ($R$), 0.028~mag ($I$), 0.022~mag ($J$), 0.020~mag ($K$).
The difference is larger for more massive MS companions and, naturally, for companions with larger physical radii (giant stars), which in extreme cases can contribute as much as 50\% to the total luminosity of the system (Karczmarek et al., in prep.).

When radial velocities are not available, we estimate upper limits to the semimajor axis and orbital periods of the Cepheid systems (see Table~\ref{CEP-orbits-noRv}). These parameters can be used to determine the feasibility of the search for companions using classical imaging, adaptive optics, or interferometry.

\subsection{PM anomalies of RR Lyrae \label{rrlyr-results}}

Table~\ref{pm_rrlyr_1} presents the results of our search for PMa in our sample of RRLs.
Out of our list of 8509 RRLs, only {\NtotalpmanomalyRRL} are present in the Hipparcos catalog  and therefore suitable for a PMa analysis.
These {\NtotalpmanomalyRRL} stars gave {five} detections with a high $\Delta>5$, and {\NmidpmanomalyRRL} stars with $3<\Delta<5$ giving a minimum binary fraction for RRLs of $P=7$\% (the known non-RRL stars were removed from this count).
In addition, {\NlowpmanomalyRRL} stars show suspected level PM anomalies at $2<\Delta<3$.

\begin{figure}
\centering
\includegraphics[width=\hsize]{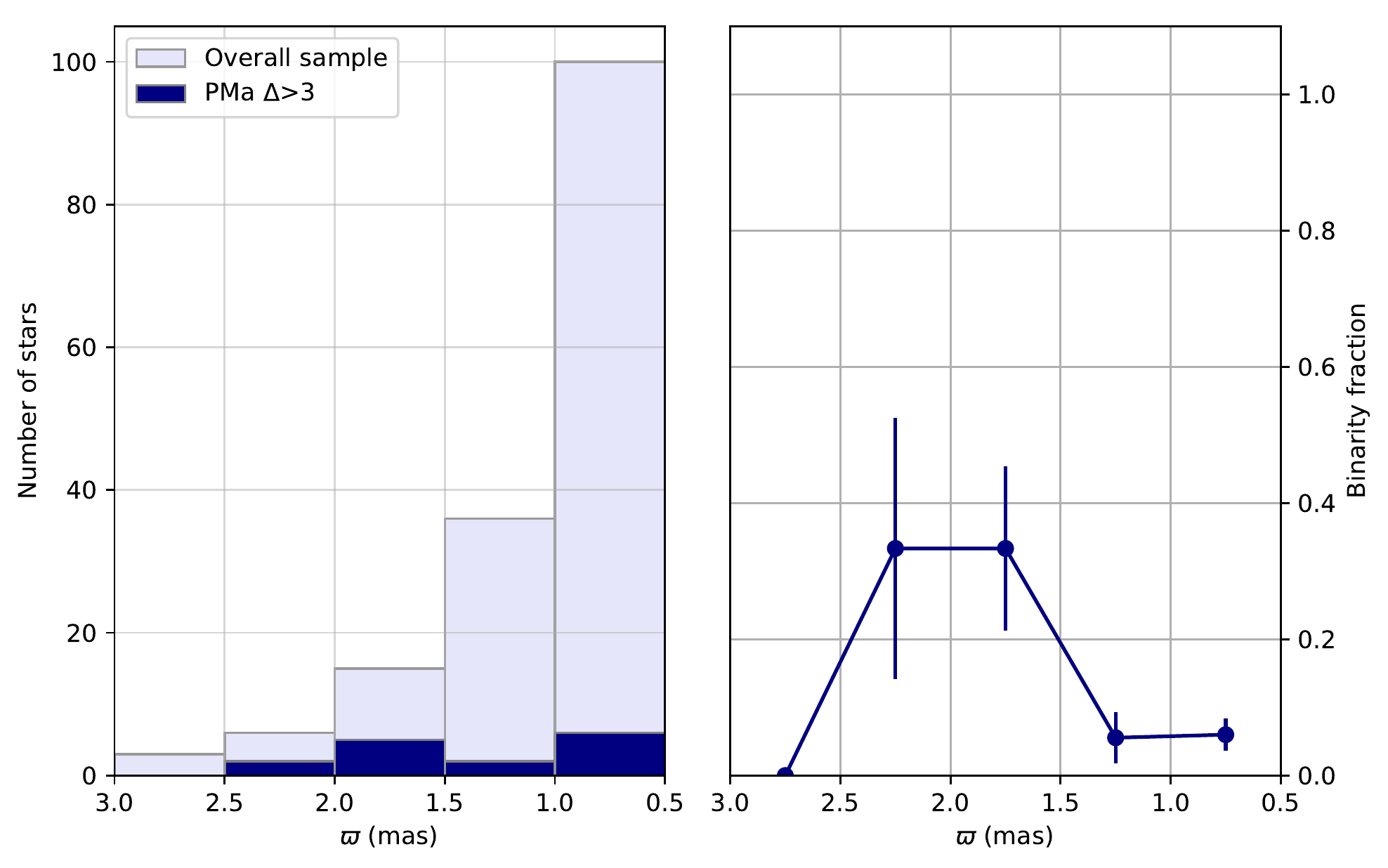}
\caption{\textit{Left:} Histogram of the RR Lyrae with detected proper motion anomalies ($\Delta>3$) as a function of parallax.
\textit{Right:} Binary fraction as a function of parallax. The error bars represent the binomial proportion 68\% confidence interval.
\label{RRL-histo}}
\end{figure}

Figure~\ref{RRL-histo} shows the statistics of the sample of RRLs with detected anomalies as a function of parallax.
We do not include in this plot the RRLs with resolved candidate companions (Paper~II).
As was true for CCs, the fraction of detected PM anomalies decreases with the parallax, due to the decreasing sensitivity of the search technique.
For the nearest targets, the binary fraction is approximately 0.4, which is probably a reasonable approximation of the true mean binary fraction of our RRL sample.

\object{TU UMa} is the only RRL that has been convincingly shown to be a member of a binary system \citepads{1986CoKon..89...57S, 1995IBVS.4205....1K, 1999AJ....118.2442W, 2016A&A...589A..94L}.
We observe a strong GDR2 PMa at $\Delta_\mathrm{G2}=6.1$ and also a significant Hipparcos PMa $\Delta_\mathrm{Hip}=2.8$ that confirms the presence of an orbiting companion.
Adopting the spectroscopic orbital parameters determined by \citetads{2016A&A...589A..94L} allows us to conduct a level 2 analysis and determine its complete orbital parameters.
The result is presented in Table~\ref{TUUMa-parameters}.
The mass that we obtain for the companion ($m_2 = 1.98 \pm 0.33\,M_\odot$) is high, and is due to the high inclination of the retrograde orbit of the system. This may imply that the companion of \object{TU UMa} is a massive white dwarf (a hypothesis already proposed by \citeads{1995IBVS.4205....1K} and \citeads{2016A&A...589A..94L}) or possibly a neutron star. Considering the old age of the RRL, a white dwarf companion will be cool and difficult to detect by imaging or interferometry, particularly as the angular separation with the primary is only on the order of 10\,mas.
It is important to note, however, that the PM vector from Hipparcos is imprecise for this star, with uncertainties larger than 1\,mas\,a$^{-1}$ on both axes. The orbital parameters will improve with the future Gaia data releases.

Estimates of the upper limits of the semimajor axes and orbital periods of a selected sample of RRLs are presented in Table~\ref{RRL-orbits-noRv}. Most of the detected RRL binaries are likely on very long-period orbits, with the exceptions of \object{AT And}, \object{CZ Lac,} and \object{AR Ser}.

Table~\ref{pm_binaries_various} presents the results of the PMa analysis for the stars that were incorrectly classified as CCs or RRLs.

\begin{table}
 \caption{Parameters of the \object{TU UMa} system from the combined analysis of the spectroscopic orbit of \citetads{2016A&A...589A..94L} (their Model 2) and the proper motion anomaly vectors.}
 \label{TUUMa-parameters}
 \centering
 \renewcommand{\arraystretch}{1.2}
 \begin{tabular}{ll}
  \hline
  \hline
  \noalign{\smallskip}
\multicolumn{2}{l}{\textit{Adopted parameters}} \\
    Parallax  from GDR2 $\varpi$         &  $1.592 \pm 0.063$\,mas \\
    Mass of RRL $m_1$     &  $0.6 \pm 0.1\,M_\odot$ \\
  \hline
  \noalign{\smallskip}
\multicolumn{2}{l}{\textit{Parameters from \citetads{2016A&A...589A..94L} }} \\
    Orbital period  $P$                   &  $8499 \pm 29$\,d \\
    Eccentricity  $e$                     &   $0.686 \pm 0.025$ \\
    Arg. of periastron  $\omega$    &     $184 \pm 2$\,deg \\
    $v_r$ amplitude   $K_1$ & $5.2 \pm 0.4$\,km\,s$^{-1}$ \\
    $v_r$ at Hip epoch         &    $+0.33 \pm 0.16$\,km\,s$^{-1}$ \\
    $v_r$ at GDR2 epoch        &    $+0.83 \pm 0.14$\,km\,s$^{-1}$ \\
  \hline
  \noalign{\smallskip}
\multicolumn{2}{l}{\textit{PMa vectors}} \\
    $\vec{\mu_\mathrm{Hip}}$  &   $[+0.2_{\pm 1.1},  +3.0_{\pm 1.1}]$\,mas\,a$^{-1}$ \\
    $\vec{\mu_\mathrm{G2}}$  &  $[+0.75_{\pm 0.12},  +0.30_{\pm 0.14}]$\,mas\,a$^{-1}$ \\
      \hline
  \noalign{\smallskip}
\multicolumn{2}{l}{\textit{Parameters from present analysis}} \\
    Inclination $i$   &  $160 \pm 6$\,deg \\
    Semimajor axis  $a$                &    $11.18 \pm 0.51$\,au \\
    Ang. semimajor axis  $\theta$   &   $17.8 \pm 1.1$\,mas \\
    Long. of asc. node  $\Omega$  &  $358 \pm 23$\,deg \\
     Mass of secondary   $m_2$ &   $1.98 \pm 0.33\,M_\odot$ \\
  \hline
\end{tabular}
\end{table}

\section{Conclusion\label{conclusion}}

We detected a significant number of new candidate companions of CCs and RRLs from the signature of their orbital motion on the proper motion vector of the targets.
CCs have long been known to have a high binary fraction, and our survey of nearby CCs indicates that their binary fraction is likely above 80\%;  in addition, there is  a significant fraction of triple or quadruple systems (e.g., Polaris, U Aql, W Sgr, AW Per, $\delta$ Cep).

The very small number of known binaries in the RRL class has long been a puzzle, but we detect significant PM anomalies $\Delta>3$ for 7\% of the 189 nearby RRLs that we surveyed, indicating that they are likely binaries. The massive companion of \object{TU UMa}, likely a white dwarf, points at the possibility that a significant fraction of RRL companions may be compact objects, which complicates their detection.

The presence of PM anomalies is an efficient way to determine the binary status of a large number of stars, and provides a valuable constraint on population synthesis models.
The most interesting candidates can easily be identified for further characterization.
In future Gaia data releases, the availability of time-resolved dynamical PM and radial velocity measurements opens the possibility, together with the parallax, of determining 3D linear velocity vectors of the photocenters of a massive number of binary and multiple systems. This will allow us to improve our understanding of their physical properties and of the role of binarity in stellar evolution.
The combination of the future Gaia data releases with targeted spectroscopic observing campaigns will enable the thorough characterization of a large number of binary and multiple stars of all types with astrophysically interesting properties, at a reasonable cost in observing time.
This will deeply improve our understanding of their physical properties and of the role of binarity in stellar evolution.

\begin{acknowledgements}
This work has made use of data from the European Space Agency (ESA) mission {\it Gaia} (\url{http://www.cosmos.esa.int/gaia}), processed by the {\it Gaia} Data Processing and Analysis Consortium (DPAC, \url{http://www.cosmos.esa.int/web/gaia/dpac/consortium}).
Funding for the DPAC has been provided by national institutions, in particular the institutions participating in the {\it Gaia} Multilateral Agreement.
The authors acknowledge the support of the French Agence Nationale de la Recherche (ANR), under grant ANR-15-CE31-0012-01 (project UnlockCepheids).
The research leading to these results  has received funding from the European Research Council (ERC) under the European Union's Horizon 2020 research and innovation program (grant agreement No 695099).
Support was provided to NRE by the Chandra X-ray Center NASA Contract NAS8-03060.
This study was funded by the NKFIH K-115709 grant of the Hungarian National Research and Innovation Office.
W.G. and G.P. gratefully acknowledge financial support for this work from the BASAL Centro de Astrofisica y Tecnologias Afines (CATA) AFB-170002. 
W.G. acknowledges financial support from the Millenium Institute of Astrophysics (MAS) of the Iniciativa Cientifica Milenio del Ministerio de Economia, Fomento y Turismo de Chile, project IC120009.
We acknowledge support from the IdP II 2015 0002 64 grant of the Polish Ministry of Science and Higher Education.
We acknowledge the past financial support to this research program of the ``Programme National de Physique Stellaire'' (PNPS) of CNRS/INSU, France.
This research made use of Astropy\footnote{Available at \url{http://www.astropy.org/}}, a community-developed core Python package for Astronomy \citepads{2013A&A...558A..33A}.
This research has made use of the International Variable Star Index (VSX) database, operated at AAVSO, Cambridge, Massachusetts, USA.
This research was supported by the Munich Institute for Astro- and Particle Physics (MIAPP) of the DFG cluster of excellence ``Origin and Structure of the Universe.''
We thank the organizers of the MIAPP workshop ``The extragalactic distance scale in the Gaia era'' held on 11 June - 6 July 2018 in Garching (Germany), which provided a place for  very useful exchanges that enhanced the work presented in the present paper.
We used the SIMBAD and VIZIER databases and catalog access tool at the CDS, Strasbourg (France), and NASA's Astrophysics Data System Bibliographic Services.
The original description of the VizieR service was published in \citetads{2000A&AS..143...23O}.
\end{acknowledgements}

\bibliographystyle{aa} 
\bibliography{../biblioCepheids}

\begin{appendix}

\section{Detected proper motion anomalies}

\begin{table*}
 \caption{Proper motion anomalies of Galactic Cepheids.
$\varpi$ is the GDR2 parallax (Sect.~\ref{gdr2corrections}), except when marked with $^*$ (Hipparcos; \citeads{2007ASSL..350.....V}) and $^+$ (\citeads{2000A&AS..143..211B}, for RY\,Vel only).
The PM vectors at the Hipparcos ($\mu_\mathrm{Hip}$) and GDR2 ($\mu_\mathrm{G2}$) epochs are compared to the mean PM computed using the Hipparcos and GDR2 positions ($\mu_\mathrm{HG}$).
The observed differences are listed in terms of signal-to-noise ratio $\Delta_\mathrm{Hip}$ and $\Delta_\mathrm{G2}$.
When the \citetads{2018A&A...616A..17A} quality parameters are not all satisfied, the star is marked with $\dag$ after $\Delta_\mathrm{G2}$ and when the RUWE $\varrho  > 1.4$ they are marked with $\ddag$.
The minimum $\Delta$ anomaly is set to S/N=5 for a strong detection ($\star$), S/N=3 for a detection ($\bullet$) and S/N=2 for a suspected binary ($\circ$).
The binary type as listed in the database by \citetads{2003IBVS.5394....1S} is provided in the ``Bin. type'' column.
The table is available at CDS.}
 \label{pm_binaries1}
 \tiny
 \centering
 \renewcommand{\arraystretch}{0.8}
\begin{tabular}{llrrcrrrrcrcc}
 \hline
 \hline
Name  & Period  & $\varpi$  & \multicolumn{2}{c}{$\mu_\mathrm{HG}$ (mas\,a$^{-1}$)} & \multicolumn{2}{c}{$\mu_\mathrm{Hip} - \mu_\mathrm{HG}$ (mas\,a$^{-1}$)} & $\Delta_\mathrm{Hip}$ & \multicolumn{2}{c}{$\mu_\mathrm{G2} - \mu_\mathrm{HG}$ (mas\,a$^{-1}$)} & $\Delta_\mathrm{G2}$ & PM & Bin. \\
 & (d) & (mas) & \multicolumn{1}{c}{$\mu_\alpha$} & \multicolumn{1}{c}{$\mu_\delta$}& \multicolumn{1}{c}{$\mu_\alpha$} & \multicolumn{1}{c}{$\mu_\delta$} & & \multicolumn{1}{c}{$\mu_\alpha$} & \multicolumn{1}{c}{$\mu_\delta$} & & bin. & type \\
  \hline \noalign{\smallskip}
\object{T Ant} &  5.90 & 0.32 & $-6.87_{\,0.03}$ & $+6.06_{\,0.04}$ & $-0.77_{\,1.02}$ & $-0.01_{\,1.05}$ &  0.8 & $-0.07_{\,0.08}$ & $-0.07_{\,0.09}$ &  1.1 &  & - \\
\object{U Aql} &  7.02 & 3.63$^*$ & $+1.00_{\,0.03}$ & $-8.97_{\,0.03}$ & $-1.99_{\,0.92}$ & $-0.17_{\,0.74}$ &  2.2 & $+1.01_{\,0.20}$ & $-3.97_{\,0.17}$ & 23.3$\,\dag$$\,\ddag$ & $\star$ & O \\
\object{SZ Aql} & 17.14 & 0.45 & $+0.32_{\,0.05}$ & $-2.67_{\,0.04}$ & $-0.10_{\,1.13}$ & $+0.31_{\,0.89}$ &  0.4 & $+0.01_{\,0.11}$ & $+0.04_{\,0.11}$ &  0.4 &  & - \\
\object{TT Aql} & 13.76 & 0.92 & $-0.43_{\,0.05}$ & $-4.68_{\,0.03}$ & $+1.60_{\,1.50}$ & $+1.35_{\,1.12}$ &  1.6 & $+0.20_{\,0.12}$ & $+0.01_{\,0.11}$ &  1.6 &  & - \\
\object{FF Aql} &  6.40 & 1.84 & $-0.14_{\,0.01}$ & $-9.34_{\,0.01}$ & $+0.58_{\,0.27}$ & $+0.36_{\,0.32}$ &  2.4 & $-0.87_{\,0.17}$ & $-0.47_{\,0.18}$ &  5.6 & $\star$ & O \\
\object{FM Aql} &  6.11 & 0.96 & $+0.31_{\,0.04}$ & $-5.40_{\,0.03}$ & $-1.93_{\,1.15}$ & $+0.02_{\,0.80}$ &  1.7 & $-0.00_{\,0.12}$ & $+0.04_{\,0.12}$ &  0.4 &  & - \\
\hline
 \end{tabular}
\tablefoot{The binary type is indicated using this code: B - spectroscopic binary (: when confirmation is needed); b - photometric companion, physical relation should be investigated; O - spectroscopic binary with known orbit; V - visual binary. }
\end{table*}

 \begin{table*}
 \caption{Continued from Table \ref{pm_binaries1}.}
 \tiny
 \label{pm_binaries2}
 \centering
 \renewcommand{\arraystretch}{0.8}
 \begin{tabular}{llrrcrrrrcrcc}
 \hline
 \hline
Name  & Period  & $\varpi$  & \multicolumn{2}{c}{$\mu_\mathrm{HG}$ (mas\,a$^{-1}$)} & \multicolumn{2}{c}{$\mu_\mathrm{Hip} - \mu_\mathrm{HG}$ (mas\,a$^{-1}$)} & $\Delta_\mathrm{Hip}$ & \multicolumn{2}{c}{$\mu_\mathrm{G2} - \mu_\mathrm{HG}$ (mas\,a$^{-1}$)} & $\Delta_\mathrm{G2}$ & PM & Bin. \\
 & (d) & (mas) & \multicolumn{1}{c}{$\mu_\alpha$} & \multicolumn{1}{c}{$\mu_\delta$}& \multicolumn{1}{c}{$\mu_\alpha$} & \multicolumn{1}{c}{$\mu_\delta$} & & \multicolumn{1}{c}{$\mu_\alpha$} & \multicolumn{1}{c}{$\mu_\delta$} & & bin. & type \\
  \hline \noalign{\smallskip}
\object{DF Cas} &  3.83 & 0.34 & $-0.26_{\,0.06}$ & $-0.18_{\,0.11}$ & $-1.15_{\,4.72}$ & $+1.86_{\,4.97}$ &  0.4 & $-0.29_{\,0.10}$ & $-0.07_{\,0.14}$ &  2.9 & $\circ$ & - \\
\object{DL Cas} &  8.00 & 0.45 & $-2.67_{\,0.02}$ & $-1.17_{\,0.03}$ & $+0.25_{\,1.08}$ & $-0.04_{\,0.76}$ &  0.2 & $+0.04_{\,0.09}$ & $-0.40_{\,0.09}$ &  4.5 & $\bullet$ & O \\
\object{DW Cas} &  5.00 & 0.38 & $-2.65_{\,0.03}$ & $-1.56_{\,0.06}$ & $-1.09_{\,1.80}$ & $+0.74_{\,1.83}$ &  0.7 & $-0.09_{\,0.09}$ & $-0.20_{\,0.10}$ &  2.3 & $\circ$ & - \\
\object{FM Cas} &  5.81 & 0.39 & $-2.09_{\,0.02}$ & $-1.58_{\,0.04}$ & $+1.40_{\,1.43}$ & $-1.70_{\,1.15}$ &  1.8 & $+0.02_{\,0.09}$ & $-0.01_{\,0.09}$ &  0.3 &  & - \\
\object{V0636 Cas} & 12.12 & 1.33 & $-0.61_{\,0.01}$ & $-0.14_{\,0.02}$ & $+0.07_{\,0.39}$ & $+0.17_{\,0.47}$ &  0.4 & $+0.07_{\,0.08}$ & $-0.07_{\,0.09}$ &  1.2 &  &  \\
\object{V Cen} &  5.49 & 1.37 & $-6.75_{\,0.02}$ & $-7.05_{\,0.02}$ & $-0.39_{\,0.93}$ & $-0.43_{\,0.66}$ &  0.8 & $+0.25_{\,0.11}$ & $+0.05_{\,0.11}$ &  2.4 & $\circ$ & - \\
\hline
 \end{tabular}
\end{table*}

 \begin{table*}
 \caption{Continued from Table \ref{pm_binaries2}.}
 \tiny
 \label{pm_binaries3}
 \centering
 \renewcommand{\arraystretch}{0.8}
 \begin{tabular}{llrrcrrrrcrcc}
 \hline
 \hline
Name  & Period  & $\varpi$  & \multicolumn{2}{c}{$\mu_\mathrm{HG}$ (mas\,a$^{-1}$)} & \multicolumn{2}{c}{$\mu_\mathrm{Hip} - \mu_\mathrm{HG}$ (mas\,a$^{-1}$)} & $\Delta_\mathrm{Hip}$ & \multicolumn{2}{c}{$\mu_\mathrm{G2} - \mu_\mathrm{HG}$ (mas\,a$^{-1}$)} & $\Delta_\mathrm{G2}$ & PM & Bin. \\
 & (d) & (mas) & \multicolumn{1}{c}{$\mu_\alpha$} & \multicolumn{1}{c}{$\mu_\delta$}& \multicolumn{1}{c}{$\mu_\alpha$} & \multicolumn{1}{c}{$\mu_\delta$} & & \multicolumn{1}{c}{$\mu_\alpha$} & \multicolumn{1}{c}{$\mu_\delta$} & & bin. & type \\
  \hline \noalign{\smallskip}
\object{RS Nor} &  6.20 & 0.45 & $-2.17_{\,0.03}$ & $-3.52_{\,0.04}$ & $+2.30_{\,1.80}$ & $+1.71_{\,1.48}$ &  1.7 & $-0.04_{\,0.10}$ & $+0.07_{\,0.10}$ &  0.8 &  & - \\
\object{SY Nor} & 12.65 & 0.43 & $-2.16_{\,0.04}$ & $-2.18_{\,0.07}$ & $-3.27_{\,2.34}$ & $+1.30_{\,2.58}$ &  1.5 & $+0.79_{\,0.10}$ & $+0.08_{\,0.11}$ &  8.0 & $\star$ & B \\
\object{TW Nor} & 10.79 & 0.38 & $-2.09_{\,0.11}$ & $-2.71_{\,0.12}$ & $-6.52_{\,5.40}$ & $-5.88_{\,4.16}$ &  1.9 & $-0.19_{\,0.15}$ & $-0.14_{\,0.16}$ &  1.5 &  & - \\
\object{GU Nor} &  3.45 & 0.58 & $-1.17_{\,0.05}$ & $-2.43_{\,0.06}$ & $-3.85_{\,2.80}$ & $-0.90_{\,2.19}$ &  1.4 & $-0.17_{\,0.10}$ & $-0.06_{\,0.10}$ &  1.7 &  & - \\
\object{Y Oph} & 17.12 & 1.39 & $-3.15_{\,0.03}$ & $-4.65_{\,0.02}$ & $+0.67_{\,0.83}$ & $+0.62_{\,0.62}$ &  1.3 & $-0.06_{\,0.15}$ & $-0.08_{\,0.13}$ &  0.7 &  & O \\
\object{BF Oph} &  4.07 & 1.21 & $+0.35_{\,0.03}$ & $-0.39_{\,0.02}$ & $-0.33_{\,1.15}$ & $+0.11_{\,0.62}$ &  0.3 & $+0.02_{\,0.13}$ & $+0.17_{\,0.10}$ &  1.7 &  & B \\
\hline
 \end{tabular}
\end{table*}

\begin{table*}
 \caption{Maximum semi-major axis and orbital period of selected binary Cepheid candidate systems without spectroscopic orbits.}
 \label{CEP-orbits-noRv}
 \tiny
 \centering
 \setlength{\tabcolsep}{5pt}
 \renewcommand{\arraystretch}{1.2}
 \begin{tabular}{lcccrrr}
\hline \hline
 Target & $\varpi$ & $m_1^\dag$ & $m_2^\dag$ & $a_\mathrm{max}$ & $a_\mathrm{max}$ & $P_\mathrm{orb\ max}$ \\
& (mas) & ($M_\odot$) & ($M_\odot$) & (lin.) & (ang.)& \\
\hline \noalign{\smallskip}
\object{RX Aur} & $ 0.57_{ 0.05}$ & $ 6.00_{ 0.90}$ & $ 3.00_{ 1.80}$ & $1.8^{+9.5}_{-1.2}$\,kau & $1.0^{+5.5}_{-0.7}$\,$\arcsec$ & $25.6^{+144.2}_{-21.1}$\,ka \\
\object{SY Aur} & $ 0.34_{ 0.05}$ & $ 5.80_{ 0.87}$ & $ 2.90_{ 1.74}$ & $0.4^{+1.4}_{-0.3}$\,kau & $0.2^{+0.5}_{-0.1}$\,$\arcsec$ & $3.2^{+20.8}_{-2.5}$\,ka \\
\object{RW Cam} & $ 1.53_{ 0.18}$ & $ 6.70_{ 1.00}$ & $ 3.35_{ 2.01}$ & $0.1^{+0.1}_{-0.0}$\,kau & $0.2^{+0.1}_{-0.1}$\,$\arcsec$ & $0.4^{+0.4}_{-0.2}$\,ka \\
\object{CK Cam} & $ 1.30_{ 0.03}$ & $ 4.10_{ 0.61}$ & $ 2.05_{ 1.23}$ & $0.4^{+0.2}_{-0.1}$\,kau & $0.5^{+0.2}_{-0.2}$\,$\arcsec$ & $3.4^{+2.0}_{-1.2}$\,ka \\
\object{RZ CMa} & $ 0.52_{ 0.04}$ & $ 4.40_{ 0.66}$ & $ 2.20_{ 1.32}$ & $0.2^{+0.1}_{-0.1}$\,kau & $0.1^{+0.1}_{-0.0}$\,$\arcsec$ & $1.2^{+1.6}_{-0.6}$\,ka \\
\object{SX Car} & $ 0.52_{ 0.04}$ & $ 4.60_{ 0.69}$ & $ 2.30_{ 1.38}$ & $1.6^{+11.0}_{-1.0}$\,kau & $0.8^{+5.7}_{-0.5}$\,$\arcsec$ & $24.3^{+235.8}_{-20.9}$\,ka \\
\hline
 \end{tabular}
\end{table*}

\begin{table*}
 \caption{Proper motion anomalies of RR Lyrae stars. The columns are identical to Table~\ref{pm_binaries1}, except ``RR'' that lists the subtype of variability.
When the \citetads{2018A&A...616A..17A} quality control parameters are not all satisfied, the star is marked with $\dag$ after $\Delta_\mathrm{G2}$ and when RUWE $\varrho > 1.4$ they are marked with $\ddag$ (Sect.~\ref{gdr2corrections}).
The table is available at the CDS.}
 \label{pm_rrlyr_1}
 \tiny
 \centering
 \renewcommand{\arraystretch}{0.8}
\begin{tabular}{llrrcrrrrcrcl}
 \hline
 \hline
Name & Period & $\varpi$ & \multicolumn{2}{c}{$\mu_\mathrm{HG}$ (mas\,a$^{-1}$)} & \multicolumn{2}{c}{$\mu_\mathrm{Hip} - \mu_\mathrm{HG}$ (mas\,a$^{-1}$)} & $\Delta_\mathrm{Hip}$ & \multicolumn{2}{c}{$\mu_\mathrm{G2} - \mu_\mathrm{HG}$ (mas\,a$^{-1}$)} & $\Delta_\mathrm{G2}$ & PM & RR \\
 & (d) & (mas) & \multicolumn{1}{c}{$\mu_\alpha$} & \multicolumn{1}{c}{$\mu_\delta$}& \multicolumn{1}{c}{$\mu_\alpha$} & \multicolumn{1}{c}{$\mu_\delta$} & & \multicolumn{1}{c}{$\mu_\alpha$} & \multicolumn{1}{c}{$\mu_\delta$} & & bin. & \\
 \hline \noalign{\smallskip}
\object{SW And} &  0.44 & 1.81 & $-6.56_{\,0.04}$ & $-18.83_{\,0.03}$ & $-0.30_{\,1.33}$ & $+1.25_{\,0.88}$ &  1.4 & $-0.20_{\,0.14}$ & $-0.56_{\,0.26}$ &  2.6 & $\circ$ & AB \\
\object{XX And} &  0.72 & 0.72 & $+58.31_{\,0.04}$ & $-32.84_{\,0.05}$ & $+1.36_{\,1.50}$ & $-1.34_{\,1.31}$ &  1.4 & $-0.31_{\,0.11}$ & $-0.13_{\,0.12}$ &  3.0 & $\bullet$ & AB \\
\object{AT And} &  0.62 & 2.21 & $-8.62_{\,0.03}$ & $-50.81_{\,0.05}$ & $+0.46_{\,1.28}$ & $+0.19_{\,1.39}$ &  0.4 & $-7.09_{\,0.45}$ & $-2.15_{\,0.39}$ & 16.7$\,\dag$$\,\ddag$ & $\star$ & AB \\
\object{WY Ant} &  0.57 & 0.94 & $+35.91_{\,0.04}$ & $-50.19_{\,0.05}$ & $+0.43_{\,1.57}$ & $+0.72_{\,1.53}$ &  0.5 & $+0.50_{\,0.17}$ & $+0.01_{\,0.15}$ &  2.9 & $\circ$ & AB \\
\object{DN Aqr} &  0.63 & 0.67 & $+46.81_{\,0.07}$ & $-15.34_{\,0.06}$ & $+3.94_{\,2.79}$ & $+3.32_{\,1.73}$ &  2.4 & $-0.09_{\,0.13}$ & $-0.20_{\,0.12}$ &  1.8 & $\circ$ & AB \\
\object{CH Aql} &  0.39 & 0.29 & $+0.77_{\,0.76}$ & $-4.97_{\,0.46}$ & $+23.44_{\,25.10}$ & $+16.55_{\,15.75}$ &  1.4 & $+0.09_{\,0.77}$ & $+0.93_{\,0.47}$ &  2.0$\,\ddag$ & $\circ$ & AB \\
\hline
 \end{tabular}
\end{table*}

\begin{table*}
 \caption{Maximum semi-major axis and orbital period of selected RR Lyrae stars. \label{RRL-orbits-noRv}}
 \tiny
 \centering
 \setlength{\tabcolsep}{5pt}
 \renewcommand{\arraystretch}{1.2}
 \begin{tabular}{lcccrrr}
\hline \hline
 Target & $\varpi^\dag$ & $m_1^\dag$ & $m_2^\dag$ & $a_\mathrm{max}$ & $a_\mathrm{max}$ & $P_\mathrm{orb\ max}$ \\
& (mas) & ($M_\odot$) & ($M_\odot$) & (lin.) & (ang.)& \\
\hline \noalign{\smallskip}
\object{XX And} & $  0.72_{  0.05}$ & $  0.60_{  0.10}$ & $  0.30_{  0.18}$ & $0.2^{+0.4}_{-0.1}$\,kau & $0.1^{+0.3}_{-0.1}$\,$\arcsec$ & $2.3^{+8.9}_{-1.7}$\,ka \\
\object{AT And} & $  2.21_{  0.27}$ & $  0.60_{  0.10}$ & $  0.30_{  0.18}$ & $3.2^{+1.0}_{-0.8}$\,au & $7.0^{+2.3}_{-1.9}$\,mas & $5.9^{+2.3}_{-1.8}$\,a \\
\object{AE Boo} & $  1.17_{  0.04}$ & $  0.60_{  0.10}$ & $  0.30_{  0.18}$ & $0.4^{+0.8}_{-0.2}$\,kau & $0.4^{+1.0}_{-0.2}$\,$\arcsec$ & $6.9^{+28.0}_{-4.7}$\,ka \\
\object{V0363 Cas} & $  0.80_{  0.03}$ & $  0.60_{  0.10}$ & $  0.30_{  0.18}$ & $0.1^{+0.1}_{-0.1}$\,kau & $0.1^{+0.1}_{-0.1}$\,$\arcsec$ & $1.5^{+2.6}_{-0.9}$\,ka \\
\object{EZ Cep} & $  0.61_{  0.02}$ & $  0.60_{  0.10}$ & $  0.30_{  0.18}$ & $0.1^{+0.9}_{-0.1}$\,kau & $61^{+574}_{-41}$\,mas & $1.1^{+8.9}_{-0.9}$\,ka \\
\object{XZ Cyg} & $  1.60_{  0.03}$ & $  0.60_{  0.10}$ & $  0.30_{  0.18}$ & $0.2^{+0.2}_{-0.1}$\,kau & $0.4^{+0.3}_{-0.2}$\,$\arcsec$ & $4.0^{+4.9}_{-1.9}$\,ka \\
\hline
 \end{tabular}
\tablefoot{$^\dag$: Gaia DR2 parallaxes (see Sect.~\ref{companion-properties}).}
\end{table*}

\begin{table*}
 \caption{Variables of various classes presenting a proper motion anomaly.
}
 \label{pm_binaries_various}
 \tiny
 \centering
 \renewcommand{\arraystretch}{0.8}
\begin{tabular}{llrrcrrrrcrcl}
 \hline
 \hline
Name & Period & $\varpi$ & \multicolumn{2}{c}{$\mu_\mathrm{HG}$ (mas\,a$^{-1}$)} & \multicolumn{2}{c}{$\mu_\mathrm{Hip} - \mu_\mathrm{HG}$ (mas\,a$^{-1}$)} & $\Delta_\mathrm{Hip}$ & \multicolumn{2}{c}{$\mu_\mathrm{G2} - \mu_\mathrm{HG}$ (mas\,a$^{-1}$)} & $\Delta_\mathrm{G2}$ & PM & Class \\
 & (d) & (mas) & \multicolumn{1}{c}{$\mu_\alpha$} & \multicolumn{1}{c}{$\mu_\delta$}& \multicolumn{1}{c}{$\mu_\alpha$} & \multicolumn{1}{c}{$\mu_\delta$} & & \multicolumn{1}{c}{$\mu_\alpha$} & \multicolumn{1}{c}{$\mu_\delta$} & & bin. & \\
 \hline \noalign{\smallskip}
\object{CW Cet} & - & 5.21 & $+65.40_{\,0.04}$ & $-19.57_{\,0.02}$ & $-1.80_{\,0.96}$ & $-0.32_{\,0.48}$ &  2.0 & $-1.99_{\,0.39}$ & $-3.82_{\,0.24}$ & 16.7$\,\ddag$ & $\star$ & Eclipsing$^a$ \\
\object{V2121 Cyg} & - & 25.60 & $+68.82_{\,0.01}$ & $+56.78_{\,0.01}$ & $-0.02_{\,0.30}$ & $+1.08_{\,0.32}$ &  3.4 & $-0.13_{\,0.13}$ & $+0.33_{\,0.15}$ &  2.5 & $\bullet$ & $\gamma$\,Dor$^{b}$ \\
\object{BX Dra} &  0.56 & 1.95 & $-10.79_{\,0.02}$ & $+8.80_{\,0.04}$ & $+2.87_{\,0.97}$ & $-0.92_{\,1.10}$ &  3.1 & $-0.13_{\,0.09}$ & $+1.28_{\,0.11}$ & 12.3 & $\star$ & Eclipsing$^c$ \\
\object{HI Dra} & - & 3.79 & $+6.59_{\,0.01}$ & $+11.68_{\,0.03}$ & $+0.88_{\,0.69}$ & $+1.93_{\,0.82}$ &  2.7 & $-0.17_{\,0.09}$ & $-0.03_{\,0.09}$ &  1.9 & $\circ$ & Ecl. or Ell.$^d$ \\
\object{HN Dra} & - & 4.29 & $+1.89_{\,0.01}$ & $+4.19_{\,0.02}$ & $+0.31_{\,0.60}$ & $+1.49_{\,0.64}$ &  2.4 & $-0.13_{\,0.09}$ & $-0.03_{\,0.10}$ &  1.4 & $\circ$ & Ell.+$\delta$\,Sct$^e$ \\
\object{S Eri} &  0.27 & 12.04 & $+39.83_{\,0.01}$ & $-87.35_{\,0.01}$ & $-0.16_{\,0.18}$ & $+0.31_{\,0.16}$ &  2.1 & $-0.39_{\,0.38}$ & $-1.07_{\,0.39}$ &  2.9 & $\circ$ & $\delta$\,Sct$^{f}$ \\
\hline
 \end{tabular}
\tablebib{
(a): \citetads{2011MNRAS.414.2602D};
(b) \citetads{2009A&A...499..967C};
(c): \citetads{2013PASJ...65....1P};
(d): \citetads{2013OEJV..160....1H};
(e): \citetads{2004yCat..34260247C};
(f): \citetads{2000A&AS..144..469R};
(g): \citetads{2011AJ....142...39H};
(h): \citetads{2003CoSka..33...38P};
(i): \citetads{1999A&A...343..202V};
(j): \citetads{2009AJ....137.3646P}.}
\end{table*}

\end{appendix}

\end{document}